\documentclass[preprint]{aastex63}

\usepackage{natbib}
\usepackage{color}
\usepackage{graphicx}
\usepackage{amsmath}
\usepackage{amssymb}
\usepackage{savesym}
\usepackage{color}
\usepackage{import}
\usepackage{tabularx}
\usepackage{textcomp}
\usepackage{verbatimbox}

\usepackage{comment}
\usepackage{todonotes}

\newcommand{\E}[1]{\times10^{#1}}

\newcommand{\U}[3]{$#1_{-#3}^{+#2}$}

\newcolumntype{Y}{>{\centering\arraybackslash}X}

\newcolumntype{M}[1]{>{\centering\arraybackslash}m{#1}}
\newcolumntype{N}{@{}m{0pt}@{}}

\shorttitle{A New Microquasar Candidate in M\,83}
\shortauthors{Soria et al.}

\begin{document}

\label{firstpage}

\title{A New Microquasar Candidate in M\,83}


\correspondingauthor{Roberto Soria}
\email{rsoria@nao.cas.cn}

\author[0000-0002-4622-796X]{Roberto Soria}
\affiliation{College of Astronomy and Space Sciences, University of the Chinese Academy of Sciences, Beijing 100049, China}
\affiliation{Sydney Institute for Astronomy, School of Physics A28, The University of Sydney, Sydney, NSW 2006, Australia}

\author{William P.~Blair}
\affiliation{Department of Physics and Astronomy, Johns Hopkins University, 3400 North Charles Street, Baltimore, MD 21218, USA}
\affiliation{Space Telescope Science Institute, 3700 San Martin Drive, Baltimore, MD 21218, USA}

\author{Knox S.~Long}
\affiliation{Space Telescope Science Institute, 3700 San Martin Drive, Baltimore, MD 21218, USA}
\affiliation{Eureka Scientific, Inc., 2452 Delmer Street, Suite 100, Oakland, CA 94602-3017, USA}

\author{Thomas D.~Russell}
\affiliation{Anton Pannekoek Institute for Astronomy, University of Amsterdam, 1098 XH Amsterdam, The Netherlands}

\author{P.~Frank Winkler}
\affiliation{Department of Physics, Middlebury College, Middlebury, VT 05753, USA}



 \begin{abstract}
 Microquasars are neutron star or black hole X-ray binaries with jets. These jets can create shock-ionized bubbles of hot plasma that can masquerade as peculiar supernova remnants (SNRs) in extragalactic surveys. To see if this is the case in the well-studied spiral galaxy M\,83, where one microquasar candidate (M\,83-MQ1) has already been identified, we studied the properties of nine SNR candidates, selected because of their elongated or irregular morphology, from the set of previously identified SNRs in that galaxy. Using multiwavelength data from {\it Chandra}, the {\it Hubble} Space Telescope, Gemini, and the Australia Telescope Compact Array, we found that at least six of our nine sources are best interpreted as SNRs. For one source, we do not have enough observational data to explain its linear morphology. Another source shows a nebular optical spectrum dominated by photo-ionization by O stars, but its excess [Fe {\footnotesize{II}}] and radio luminosity suggest a possible hidden SNR. Finally, one source (S2) shows an elongated structure of ionized gas, two radio sources along that line, and an accretion-powered X-ray source in between them (the {\it Chandra} source L14-139). While S2 could be a chance alignment of multiple SNRs and one X-ray binary, it seems more likely that it is a single physical structure powered by the jet from the accreting compact object. In the latter case, the size and luminosity of S2 suggest a kinetic power of  $\sim$10$^{40}$ erg s$^{-1}$, in the same class as the most powerful microquasars in the local universe ({\it e.g.}, S26 in NGC\,7793 and SS\,433 in our own Galaxy).
\end{abstract}


\section{Introduction} 
\label{intro}
Jets and collimated outflows are increasingly recognized as a fundamental component of energy transport in accretion processes at all scales, from AGN and quasars \citep{2012ARA&A..50..455F,2012Sci...338.1445N} to stellar-mass compact objects \citep{2014SSRv..183..323F}, from tidal disruption events \citep{2011Natur.476..421B} to kilonovae \citep{2013Natur.500..547T} and Gamma-ray bursts \citep{2015PhR...561....1K}. The mechanical power carried by jets can alter the ambient medium on scales of hundreds of pc around stellar-mass compact objects, or hundreds of kpc around supermassive black holes. Mechanical power generated by accretion may have blown away the gas and quenched star formation in the early phases of evolution of the most massive galaxies \citep{2008MNRAS.391..481S}; it heats the diffuse gas in elliptical galaxies, groups and clusters \citep{2012NJPh...14e5023M}; more locally, it creates large bubbles of ionized gas around ultraluminous X-ray sources \citep{2008AIPC.1010..303P}.

Supernova remnants (SNRs) and, more generally, radio/optical bubbles classified as candidate SNRs in extragalactic surveys are one class of astrophysical sources where the presence and effect of jets has only recently begun to be noticed and investigated. 
Morphological studies of young SNRs in the Milky Way and Magellanic Clouds have shown  that $\approx$30--40\% of them have ``ears", defined as a pair of protrusions or lobes sticking out of the main SNR shell in opposite directions \citep{2017MNRAS.468.1226G}. (This estimate takes into account the number of cases that have actually been observed, corrected for the additional unobservable fraction of cases in which the ears would be pointing along our line of sight). Such features are usually telltale signs of currently or recently active jets. The kinetic energy of the gas in the ears is $\approx 5$--$15\%$ of the total kinetic energy of the SNR shell \citep{2017MNRAS.468.1226G}. 

There are several possible origins for such protrusions, depending on whether jets operated before, during, or after the core-collapse event. The first possibility is that the circumstellar medium (CSM) before the explosion already contained an elongated lower-density cavity or lobes, produced by jets from a high-mass X-ray binary (consisting of a neutron star or black hole receiving gas from a massive donor star) before the core collapse of the donor star in the system. The second possibility is that a substantial fraction of SNe produce jets. It is already well established that Gamma-ray bursts are associated with Type-Ic SNe \citep[{\it e.g.},][]{2009MNRAS.396.2038B,2008Sci...321.1185M,2005ApJ...620..355L,2003ApJ...599..408B,2003MNRAS.345..575M,2001ApJ...550..410M,1999ApJ...516..788W}. Other types of SNe may also produce jets \citep{2019ApJ...878...24G}.  A third possibility is that some shock-ionized bubbles in young environments of nearby galaxies have been mistakenly classified as SNRs, when they are in fact entirely generated by jets and winds of an accreting compact object ({\it e.g.}, S26 in NGC\,7793: \citealt{2010Natur.466..209P}); we will use the term ``microquasar" for this type of system, powered either by a stellar-mass black hole or by a neutron star. 

Some systems may display a combination of SNR and microquasar features: that is, the jet-like features are imprinted onto an SNR shell at later stages, inflated by the jets of an X-ray binary or a pulsar formed after the explosion. For example, this is the most common interpretation for the Galactic source SS\,433/W50, in which the jet lobes powered by the microquasar SS\,433 stick out of the W50 SNR \citep{1998AJ....116.1842D,2004ASPRv..12....1F,2007A&A...463..611B,2011MNRAS.414.2838G,2017MNRAS.467.4777F}. 

Finally, the irregular, bilateral, or elongated morphology of some SNRs may be due entirely to the combined effects of stellar winds and proper motion of the progenitor, which compress the circumstellar gas and magnetic field preferentially in one direction, producing a stellar bow shock. Three-dimensional magneto-hydrodynamical simulations predict  that this compression effect will then determine the morphology and radiative properties of the subsequent SN shock wave \citep{2018ApJ...867...61Z}. 

Distinguishing between normal SNRs that expand in an inhomogeneous medium, elongated SNRs caused by jetted SNe, and shock-ionized microquasar bubbles provides important tests for our theoretical understanding of SNe and of accretion physics. A census of jet signatures in young Galactic SNRs and shock-ionized nebulae is hampered by the limited number of such sources in the Milky Way, and our edge-on view of the Galactic disk.  At the other extreme, lack of spatial resolution in galaxies at $\ga 10$ Mpc hampers imaging studies of candidate SNRs. 

The grand-design spiral galaxy M\,83 is one of the best targets for morphological studies of young SNRs. Located a distance of 4.6 Mpc \citep{saha06} (angular scale: 1$^{\prime\prime }\approx 22$ pc), it is almost face-on ($i \approx 24^{\circ}$: \citealt{talbot79}) and it lies along a line of sight with very low optical extinction ($A_V \approx 0.18$ mag, corresponding to an absorbing column density $N_{\rm H} \approx 4 \times 10^{20}$ cm$^{-2}$: \citealt{schlafly11,kalberla05}). Its high star formation rate (between 3 and 5 $M_{\odot}$ yr$^{-1}$: \citealt{boissier05}) has created a wealth of young SNRs, with six SNe observed in the last century (\citealt{stockdale06} and references therein). 

We are conducting a wide-ranging investigation of the stellar birth and death cycle in M\,83, using X-ray, optical and radio observations (as detailed in Section 2). More than 300 SNRs and SNR candidates have been identified in our survey, from narrow-band imaging studies with the Magellan Telescope \citep{2012ApJS..203....8B} and with the {\it Hubble} Space Telescope ({\it HST}) \citep{2010ApJ...710..964D,2014ApJ...788...55B}. Follow-up spectroscopic studies of many of those candidates were carried out with the Gemini telescope \cite{2017ApJ...839...83W}. See also \cite{williams19} for the most updated list of SNRs in M\,83 and the properties of their stellar progenitors. About 1/3 of the optical SNRs have been detected with the {\it Chandra} X-ray Observatory \citep{2014ApJS..212...21L}, and about 1/4 have a radio counterpart detected with the Australia Telescope Compact Array (ATCA) (\citealt{2014ApJS..212...21L}; T.~D.~Russell et al., in prep.).

In the course of searching for SNRs in M\,83, we  \citep{2014Sci...343.1330S} found an object, hereafter MQ1, with exceptionally strong radio luminosity (flux density of $\approx 1.8$ mJy at 5.5 GHz, corresponding to a luminosity of $\approx$2 $\times 10^{35}$ erg s$^{-1}$), and elevated [S~{\footnotesize{II}}]/H$\alpha$ line ratio, associated with a bipolar structure (core and two lobes). The X-ray properties of the core (in particular, its disk-blackbody spectrum and the correlated variability between luminosity and peak disk temperature) are consistent with an accreting compact object. Thus, we argued that MQ1 is a candidate microquasar, in the same class as SS\,433 and NGC\,7793-S26. Based on the optical line luminosity, we estimated a mechanical power $P_{\rm kin} \approx 10^{40}$ erg s$^{-1}$.

Here we present a detailed multiband study of a sample of SNR candidates in M\,83, selected for their peculiar (irregular or elongated) morphology. We  also investigate whether any of those sources is a microquasar bubble, by analogy with the properties of MQ1. Our paper is organized as follows: In Section \ref{sec:methods}, we describe the archival and new data we use for this study and the selection of nine SNR candidates whose properties we analyze in the remainder of the paper.  In Section \ref{sec:results}, we present our results, summarizing the multi-band properties of the various candidates. We also update the properties of MQ1, taking into account data that were obtained since the publication of \cite{2014Sci...343.1330S}.  In Section \ref{sec:discussion}, we discuss the results, concluding that we have found a second microquasar candidate, and that one of the other objects is probably neither an SNR nor a microquasar but, instead, a particularly luminous \ion{H}{2} region.  Finally, in Section \ref{sec:conclusions}, we summarize our conclusions.



\section{Methods} 
\label{sec:methods}

\subsection{Observations used in this study}
\label{sec:observations}



{\it UV/optical/infrared bands.} \cite{2014ApJ...788...55B} and \cite{2010ApJ...710..964D} carried out {\it HST} Wide Field Camera 3 (WFC3) imaging to search for SNRs in M\,83. The galaxy was covered by a mosaic of seven WFC3 fields, which were imaged in nine narrow-band and broad-band filters. The first two fields (including the nuclear region) were observed in 2009 August and 2010 March (ID 11360; PI: R.~O'Connell) and the other five were observed from 2012 July to 2012 September (ID 12513; PI: W.~P.~Blair). We refer the readers to \citep{2014ApJ...788...55B} for the details on the WFC3 observational set-up ({\it e.g.}, orientation of the seven fields, set of filters used for the survey, and exposure times).  The {\it HST} imaging survey was combined with a ground-based imaging survey \citep{2012ApJS..203....8B} with the Inamori Magellan Areal Camera and Spectrograph (IMACS) on the 6.5-m Magellan telescope; and by 
follow-up spectrophotometric observations \citep{2017ApJ...839...83W} with the Gemini Multiobject Spectrograph (GMOS) on the 8.2-m Gemini-South telescope.

{\it X-ray band.} We observed M\,83 with the Advanced CCD Imaging Spectrometer (ACIS) on board the {\it Chandra} X-ray Observatory, with ten visits spaced between 2010 December 23 and 2011 December 28 (Observing Cycle 12), for a total of 729 ks; in all those observations, the target galaxy was placed on the back-illuminated ACIS-S3 chip. We refer to \cite{2014ApJS..212...21L} for a comprehensive discussion of the {\it Chandra}/ACIS setup and its astrometric calibration. In addition, we integrated the 2010--2011 dataset with the ACIS-S observations of 2000 April 29 (51 ks; Cycle 1) and 2001 August 4 (10 ks; Cycle 2), which provide higher sensitivity to the softest energies.


{\it Radio band.} We mapped M\,83 with the ATCA, with three sets of observations: with 3$\times$12 hours in 2011, 2015, and 2017. The data were recorded simultaneously at central frequencies of 5.5 and 9.0 GHz, with a bandwidth of 2 GHz at each frequency band. The 2011 and 2015 observations were taken with the telescope in its most extended 6 km configuration, while the 2017 observations were taken with the telescope in its more compact 1.5 km configuration. The more compact configuration provided more short baselines to increase the sensitivity of our final radio map to diffuse emission. See \cite{2014ApJS..212...21L} for a preliminary report on the results based on the 2011 ATCA observations alone. A full presentation of the radio results is in preparation (T.~D.~Russell et al., in prep).


\subsection{Sample selection}

To create a list of objects for study, we inspected the images of SNRs in the {\it HST} data, looking for sources with distinct bipolar shape, hot spots, or other unusual asymmetrical morphology.  At 4.6 Mpc, the 0\arcsec.04 pixel scale of the WFC3 UVIS camera corresponds to just over 1 pc, providing excellent data for assessing the morphology of even the smallest objects expected to be of relevance.  Based on this inspection, we selected nine objects for more detailed study, as listed in Table 1.  This list is not necessarily exhaustive, as there are a number of other objects within the total SNR candidate sample in M\,83 that might also be considered as showing interesting and unusual morphologies. Here, we have chosen the clearest examples of SNRs and SNR candidates with peculiar morphologies in an attempt to understand the physical nature of such objects. 

The list includes the two sources that were flagged as ``worthy of special mention" by \citet[][their Section 3.1 and Table 2]{2010ApJ...710..964D} because of their jet-like or bipolar morphology. Those two sources are listed as S5 and S6 in our Table 1 (presented in order of Right Ascension). 
Further out along the spiral arms, we identified two strong line emitters (S2 and S7 in our Table 1), with interesting elongated morphologies that appear to extend over $\sim$100 pc in length. We then selected five candidate SNRs with irregular morphology (S1, S3, S4, S8, and S9). 
Finally, for comparison, we also included the microquasar MQ1 in the Table, which we re-analyzed  in the same way as for the other nine sources.  
The spatial distribution of the selected objects projected on an H$\alpha$ image of M\,83 is shown in Figure 1. Narrow-band and broad-band {\it HST} images of the fields around each object are shown in Figures 2--4.  Seven of the nine candidate SNRs have been detected in X-rays; those without a {\it Chandra} detection are S6 and S7.

It is worth noting that what we define here as source S2 was identified by \cite{2012ApJS..203....8B} as two separate SNR candidates, B12-96 and B12-98. However, there is an elongated line-emission feature that connects the two optical sources (Figure 2), and there is a single, point-like {\it Chandra} counterpart (L14-139) located roughly in the middle between the two optical peaks. Therefore, for the purpose of this study, we consider the whole complex as a single astrophysical object with a peculiar, extended morphology.

It is also useful to clarify our definition of S7 (discussed in more details in Section 3.1). The field around this source contains a blue and a red star cluster separated by about 20 pc (Figure 4) along a north-south axis. The strong optical/IR line emission and radio emission coincide with the reddened star cluster, and that is what we define as the S7 source in our study. There is no X-ray source associated with the reddened star cluster; there is, instead, a {\it Chandra} source (consistent with an ordinary SNR) at the location of the blue cluster (dashed error circle in Figure 4) but that is not the object of this study.



\subsection{Data Analysis}

{\it HST}. Linear combinations of WFC3 broad-band filter images were used to create rescaled continua, which were subtracted from the corresponding narrow-band images, as described in \cite{2014ApJ...788...55B}. For this work, we used the continuum-subtracted images already prepared for that earlier study. 
We used the imaging and photometry tool {\sc {ds9}}, to select source and background regions for our target objects, and extract the instrumental count rates. We used the PHOTFLAM parameter for the relevant filters in WFC3-UVIS and WFC3-IR detectors to convert from count rates to flux densities, and the filter widths listed in the WFC3 Instrument Handbook to convert to total fluxes in each (narrow) band.

{\it Chandra}. We downloaded and reprocessed the data from the {\it Chandra} archives, for all the ACIS observations.
We used standard tasks within the Chandra Interactive Analysis of Observations ({\sc CIAO}) version 4.10 \citep{2006SPIE.6270E..1VF}. First, we reprocessed the data with {\it chandra\_repro}. Then, we applied the {\it merge\_obs} task, which reprojects all the datasets to the same tangent point and merges them, creating a coadded, exposure-corrected image. Filtered images in different energy bands were obtained with the task {\it dmcopy}. We used {\it specextract} to build spectra and response files individually from each observation and combine them into a merged spectrum, because the X-ray flux of typical SNRs is too faint for meaningful spectral analysis of individual observations.

{\it ATCA}.  The ATCA radio observations were flagged and calibrated following standard procedures within MIRIAD \citep{1995ASPC...77..433S}. Flux and phase calibration were done using PKS~1934$-$638 and 1313-333, respectively. The data were then imaged within CASA \citep{2007ASPC..376..127M}. To help minimize sidelobes from the bright central region, images were created with a Briggs robust parameter of -1. Individual images were created at each central frequency (5.5 and 9\,GHz), where data from each epoch were stacked to provide the highest sensitivity at each observing band. We also created a single image including both frequency bands from all epochs to increase sensitivity (T.~D.~Russell et al., in prep.).


{\it Gemini spectroscopy}. Eight of the nine SNRs candidates studied in this work were observed with the GMOS spectrograph. We used the results published by \cite{2017ApJ...839...83W}. 
In the context of our present work, we obtain three important pieces of information from the Gemini results, which enhance and integrate the information provided by the {\it HST} images: i) an estimate on the optical extinction, from the observed ratio of H$\alpha$ and H$\beta$ fluxes (assuming a theoretical intrinsic ratio of 2.86); ii) a means to distinguish the fractional contributions of H$\alpha$ and [N~{\footnotesize{II}}]$\lambda\lambda$6548,6584 to the combined emission imaged by the {\it HST}/WFC3-UVIS camera in the F657N filter; iii) a constraint on the electron density of the ionized gas, from the flux ratio of the [S~{\footnotesize{II}}]$\lambda\lambda$6716,6731 doublet.



\section{Results}
\label{sec:results}

\subsection{Optical Line Fluxes and Luminosities} 

We measured the observed continuum-subtracted optical fluxes in the four {\it HST}/WFC3 narrow-band filters: F502N, F657N, F673N and F164N (Table 2). 
For all sources except S7, the flux ratio between the F673N and F657N bands (proxy for the standard diagnostic ratio of [S {\footnotesize{II}}]$\lambda\lambda$6716,6731/H$\alpha$) is between $\approx$ 0.2--0.5. S7 stands out with a much lower ratio of $\approx$0.07. This is consistent with the Gemini/GMOS spectral classification of S7 as a photo-ionized H~{\footnotesize{II}} region rather than a shock-ionized SNR \citep{2017ApJ...839...83W}.

The F502N, F673N and F164N bands are dominated by [O {\footnotesize{III}}]$\lambda$5007, [S {\footnotesize{II}}]$\lambda\lambda$6716,6731, and [Fe {\footnotesize{II}}]$\lambda$1.64$\mu$m, respectively.
Determining the H$\alpha$ luminosity from F657N is approximate, because that filter is also passes [N {\footnotesize{II}}]$\lambda\lambda$6548,6584. Whenever possible ({\it i.e.}, for 8 of our 9 targets, see Table 3), the relative contributions from H$\alpha$ and [N {\footnotesize{II}}] can be directly measured from the Gemini/GMOS spectra and the actual H$\alpha$ intensity determined  \citep{2017ApJ...839...83W}. For the shock-ionized objects, the H$\alpha$ contribution varies between 0.23 and 0.53 times the flux in the F657N band; the average of those values is $\approx$0.40. A larger contribution of 0.59 is seen in S7: this is again consistent with its classification as a photo-ionized H~{\footnotesize{II}} region (although this value by itself does not rule out a shock-ionized SNR buried within the bright photoionized region). 

For the only target that does not have a Gemini/GMOS spectrum (S5), we assumed the average ratio of $F({\rm {H}}\alpha) = 0.4 F({\rm{F657N}})$ derived above from the shock-ionized sources. Looking at the whole sample of about 140 SNR candidates with Gemini spectra, it is clear (\citealt{2017ApJ...839...83W}, their Figs.~8 and 9) that there is a large spread in the [N {\footnotesize{II}}]/H$\alpha$ flux ratios, corresponding to $F({\rm {H}}\alpha)/F({\rm{F657N}})$ spanning the whole range between $\sim 0.25$ to  0.7, possibly as a function of shock velocity and local metal abundance. Thus, our choice of $F({\rm {H}}\alpha) = 0.4 F({\rm{F657N}})$ is roughly the mid-point (on a log scale) of that range.
(For comparison, \citealt{2014ApJ...788...55B} assumed a slightly lower value of $F({\rm {H}}\alpha) = 0.33 F({\rm{F657N}})$ for sources without direct line-ratio measurements, and $F({\rm {H}}\alpha) = 0.5\, F({\rm{F657N}})$ was assumed by \citealt{2014Sci...343.1330S} for MQ1.)  

The optical spectra also provide direct information on the density of the emitting plasma, from the observed flux ratio of the sulfur doublet ([S {\footnotesize{II}}]$\lambda6716$/[S {\footnotesize{II}}]$\lambda6731$). The WFC3 filter F673N includes both lines, but the ratio comes from the Gemini spectra. We find (Table 3) a range of values between $\approx0.9$ and $\approx$1.35, which correspond to electron densities between $\approx$70---800 cm$^{-3}$, assuming a temperature near $\sim$10$^4$ K for the S${^+}$ zone \citep{2016ApJ...816...23S,of2006}.


Next, we corrected the observed fluxes for the effect of optical extinction. In addition to the Galactic line-of-sight value ($A_V \approx 0.18$ mag), we need to take into account the local extinction in the star-forming disk of M\,83. For the eight objects with Gemini spectra, 
we determined how the observed ratio between the H$\alpha$ and H$\beta$ line fluxes differs from the canonical value of 2.86; we assumed that the reduced contribution from H$\beta$ is due to extinction ($A_{\rm{H}\alpha} \approx 0.818 A_V$; $A_{\rm{H}\beta} \approx  1.164 A_V$). The observed H$\alpha$/H$\beta$ flux ratios span the range  $\approx 3.7$ to 6.7 (Table 3). This corresponds to total extinctions (line-of-sight plus intrinsic) $A_V$ from  $\approx0.8$ mag (for S9) to $\approx2.7$ mag (for S2). We used the average of those eight values, $\langle A_V \rangle = 1.7$ mag, as a plausible estimate for the unknown extinction of S5. 
Putting together the previous findings, and scaling to the distance of 4.6 Mpc, we determined the intrinsic luminosities (Table 4) in the [O~{\footnotesize{III}}], H$\alpha$, and [Fe {\footnotesize{II}}]$\lambda$1.64$\mu$m lines (for [Fe {\footnotesize{II}}],  in only eight of the nine targets). For comparison, we also re-calculated the line emission from MQ1 (Table 4) using the same procedure as for the nine SNR candidates. For MQ1, we adopted a total extinction $A_V = 3.9$ mag, as discussed in \cite{2014Sci...343.1330S}. The higher extinction for MQ1 is not surprising, given its proximity to the dusty nuclear starburst region.

[Fe {\footnotesize{II}}]$\lambda$1.64$\mu$m is often used as a tracer of cooler gas in star-forming regions and starburst galaxies ({\it {e.g.}}, \citealt{2006ApJS..166..188L,2003AJ....125.1210A,1989A&A...214..307O,2000ApJ...528..186M}). Fe$^{+}$ has an ionization potential of only 16.2 eV; therefore, it is easily ionized to Fe$^{++}$ in H {\footnotesize{II}} regions, with the result that the [Fe {\footnotesize{II}}] line emission from photoionized gas is strongly suppressed. Instead, Fe$^{+}$ survives in cooler, partially ionized and recombining regions, such as the cooling flows behind SNR shocks. In particular, [Fe {\footnotesize{II}}] traces dense, collisionally excited gas at temperatures of $\approx$6,000 K, and partially ionized, X-ray heated gas at temperatures $\approx$8,000 K \citep{2000ApJ...528..186M}.
If we assume that most of the [Fe {\footnotesize{II}}] emission comes from the SNR recombination zones, we can use the dereddened flux ratio between [Fe {\footnotesize{II}}]$\lambda$1.64$\mu$m and H$\beta$ as an independent check of our estimated values of extinction and H$\alpha$/[N {\footnotesize{II}}] flux ratios. 


%

%



Assuming the canonical value $L_{\rm{H}\alpha} \approx 2.87 L_{\rm{H}\beta}$, we infer from Table 4 that $L_{\rm{[Fe~II]}}/L_{\rm{H}\beta} \approx 0.20$--0.45 for seven (specifically, S1, S3, S4, S5, S6, S8, S9) of the eight SNR candidates for which an [Fe {\footnotesize{II}}] measurement is available. This range of values is consistent with the model predictions \citep{2008ApJS..178...20A} for shock-ionized gas with solar abundances, interstellar medium (ISM) densities in the range $n_e \sim$1--10 cm$^{-3}$, equipartition magnetic field, and shock velocities $\approx$100--500 km s$^{-1}$. It is also consistent with the values observed from nearby SNRs \citep{1989A&A...214..307O}. 

The only outlier in our sample is again the S7 nebula, which has $L_{\rm{[Fe~II]}}/L_{\rm{H}\beta} \approx 0.03$ (Table 4). This low value is inconsistent with any shock velocity $\gtrsim$100 km s$^{-1}$ \citep{2008ApJS..178...20A}, but is consistent with the interpretation of S7 as an H {\footnotesize{II}} region. Optical and infrared [Fe {\footnotesize{II}}] emission lines are sometimes seen in H {\footnotesize{II}} regions, for example in the Orion nebula \citep{1992ApJ...389..305O,1998ApJ...492..650B,1999A&A...348..222R}. The infrared [Fe {\footnotesize{II}}] lines are generally attributed to collisional excitation in partially ionized zones close to the ionization front, at densities $n_e \sim 10^2$--$10^4$ cm$^{-3}$ \citep{2000ApJ...543..831V,1998A&A...330..696M,1998ApJ...492..650B}.

Given the exceptionally strong H$\alpha$ (Table 4) and radio (Table 2) luminosity of the S7 nebula, and its peculiar properties compared with the rest of the sample, we investigated this source further. The broad-band optical image (Figure 4) of the S7 field shows a highly reddened cluster of massive stars (identified as cluster NGC\,5236-2-530 in the catalog of \citealt{2011A&A...529A..25S}), coincident with the peak of the H$\alpha$ and [Fe {\footnotesize{II}}] emission, as well as with the position of the unresolved radio emission (discussed later in Section 3.3); it also shows a bright cluster of blue stars located $\sim$20 pc to the north of the peak radio emission. A soft, faint ($L_{\rm X} \approx 10^{36}$ erg s$^{-1}$) thermal X-ray source (L14-275: \citealt{2014ApJS..212...21L}) is centered at the location of the blue stars rather than at the position of the optical and radio nebula (Figure 4). From aperture photometry on the broad-band {\it HST} images, we estimate that the integrated (and de-reddened) absolute brightness of the cluster at the central position of S7 is $M_V \approx -9.7$ mag, $M_B \approx-10.0$ mag. Using the stellar population code {\sc starburst99} \citep{1999ApJS..123....3L}, we find that this is consistent with an instantaneous burst of star formation, at an age $\lesssim$3 Myr and a stellar mass $\approx$(1.0--1.5) $\times 10^4 M_{\odot}$. Such a population is expected to produce an ionizing photon flux $Q(\rm{H}0) \approx$(3.0--4.0) $\times 10^{50}$ photons s$^{-1}$ above 13.6 eV, mostly from main-sequence O stars. 

On the other hand, we have already measured the intrinsic H$\alpha$ luminosity of S7 (Table 4), and we can infer its H$\beta$ luminosity by applying the standard Balmer decrement of 2.87 for photoionized gas at 10,000 K \citep{of2006}; we obtain $L_{\rm{H}\beta} \approx 1.5 \times 10^{38}$ erg s$^{-1}$. We also know from atomic physics that the emission of one H$\beta$ photon is triggered every $\sim$8.5 ionizing photons between 13.6 eV and $\sim$0.2 keV \citep{of2006}. Hence, the  observed  Balmer-line emission implies  an  ionizing  photon  flux $Q(\rm{H}0) \approx 3.1 \times 10^{50}$ photons s$^{-1}$, in good agreement with the brightness of the optical continuum. Interestingly, the other cluster of blue stars to the northeast of S7 is brighter in the blue band because it is dominated by blue supergiants at an age of $\sim$10 Myr, but it has an ionizing flux an order of magnitude lower than the highly reddened cluster coincident with S7.

It is interesting to compare our results (mostly based on line emission) with the findings of \cite{williams19}, who fitted the {\it HST}/WFC3 broad-band colours of the stars within a 50 pc radius of most of the SNR candidates in M\,83. From that modelling, \cite{williams19} determined the age distribution of the stellar populations and the corresponding zero-age main-sequence mass of the most massive stars currently present around each SNR (a proxy for the progenitor mass of the SNR). For eight of the nine SNR candidates in our target list (all except S7), the stellar progenitor mass spans a range between $\approx$8 M$_{\odot}$ (for S8) and $\approx$17 M$_{\odot}$ (for S3), which is the typical mass range of SN progenitors \citep{williams19}. By contrast, they find a zero-age main-sequence mass $\approx$40 M$_{\odot}$ for the most massive stars around S7 (region 207 in their catalog). This is consistent with our interpretation that the S7 optical nebula is mainly the result of photoionization by O stars. Note that the 50-pc source region used by \citep{williams19} includes not only the reddened star cluster but also most of the blue cluster.

\subsection{X-ray Colors and Spectra}

Seven of the nine candidate SNRs selected for this study have  point-like X-ray counterparts (Table 1). Using the {\sc ciao} task {\it srcflux}, we estimated and compared the count rates and fluxes of each source in each of the 12 {\it Chandra}/ACIS-S observations available in the archive, searching for variability between observations (the count rates are not high enough to constrain intra-observational variability). X-ray variability by an order of magnitude \citep{2014Sci...343.1330S} was one of the main clues that suggested a microquasar interpretation for MQ1 ({\it Chandra} source L14-237). We find significant variability only in one SNR candidate: L14-139, counterpart of S2.
The variability of L14-139 is less dramatic than that of MQ1; it spans a factor of a few in flux over the whole series of observations, with no clear systematic trend. In Table~\ref{tab:S2_flux_table}, we report the model-independent, observed fluxes in the 0.5--7 keV band (``broad'' band in {\it srcflux}) for each of the 12 observations, and a model-dependent flux in the same band, based on an absorbed power law with photon index $\Gamma = 1.5$ and total absorbing column $N_{\rm H} = 10^{21}$ cm$^{-2}$\footnote{The choice of this simple model as input of {\it srcflux} is justified by the color and spectral analysis discussed later in this section. For this model, the observed flux in the 0.3--10 keV band is $\approx$1.3 times higher than that in the 0.5--7 keV band.} 
All other sources are consistent with constant fluxes throughout the 12 observations, as we expect from hot plasma in SNRs.  The lack of an X-ray counterpart for the S7 nebula, down to a detection limit of $\approx$5 $\times 10^{35}$ erg s$^{-1}$, supports the idea that S7 is an H {\footnotesize{II}} region, photoionized by stellar sources. 

%

The second step of our X-ray analysis is an assessment of the average source colours. We used the soft (0.35--1.1 keV), medium (1.1--2.6 keV) and hard (2.6--8.0 keV) photon fluxes already calculated by \cite{2014ApJS..212...21L}, and plotted them in the traditional color-color diagram often used for X-ray source classifications \citep{2003ApJ...595..719P}. We find (Figure 6) that six of the sources are located in the band usually occupied by SNRs, or more generally by any source consisting of an optically thin thermal-plasma emission at (average) temperatures $\sim$0.5 keV; see, {\it e.g.}, \citealt{2014ApJS..212...21L,2003A&A...410...53S,2003ApJ...595..719P} for other examples of the source classification in X-ray color-color diagrams of M\,83. Instead, L14-139 (S2) stands out for its harder colors, as does L14-237 = MQ1, just as both sources also stand out for variability. L14-139 is located in the sector of the diagram occupied by power-law sources with a hard spectrum (photon index $\Gamma \sim 1.5$). L14-237 is located in the region of the diagram usually populated by X-ray binaries in the high/soft state (steep power-law or disk-blackbody); this is confirmed by detailed spectral analysis  \citep{2014Sci...343.1330S}.

The third step is spectral modelling of the stacked spectra from all {\it Chandra}/ACIS-S observations, for each source. As expected from our color-color analysis, six sources (L14-63 = S1, L14-183 = S3, L14-186 = S4, L14-256 = S5, L14-310 = S9, and L14-358 = S9) have spectra consistent with thermal plasma emission ({\it mekal} model in {\sc xspec}). In most cases (Table 6, and Figures 7, 8), two temperature components are needed: one component with a temperature $kT_1 \lesssim 0.3$ keV and the other with a temperature $kT_2 \sim 0.5$--1.0 keV; only the lower-temperature component is statistically needed for the spectrum of L14-63 (S1). 

The only source that stands out is L14-139 (S2), as already suspected from its X-ray colors. Spectral analysis confirms that L14-139 has a power-law dominated spectrum, with a photon index $\Gamma = 1.3 \pm 0.3$, plus a soft excess consistent with thermal emission (Table 6). The soft thermal component contributes for $\approx$13\% of the observed flux, and 28\% of the de-absorbed luminosity. The hard non-thermal spectrum supports the interpretation of this source as an X-ray binary. In most of the 12 {\it Chandra} observations (in particular, in all the longer ones), the source was located several arcmin away from the aimpoint on the ACIS chip, to the effect that the point spread function was somewhat degraded; thus, we are unable to determine whether the soft emission is more spatially extended than the hard component. We can only say that both the soft and the hard emission are consistent with a point-like source at that location, with a full width at half maximum $\lesssim$50 pc.

In the optical continuum, L14-139 has a point-like counterpart: a bright, red star with apparent magnitude $m_{555W} = 25.69 \pm 0.05$mag, $m_{814W} = 21.46 \pm 0.05$ mag. Assuming the same extinction $A_V = 2.66$ mag inferred for the ionized nebula (Table 3),
we obtain a de-reddened absolute brightness $M_V \approx -6.5$ mag, $M_I \approx -8.4$ mag, typical of a red supergiant (Figure 2). A somewhat different value of the extinction can be obtained from the best-fitting column density $N_{\rm H}$ (Table 6), converted to an optical extinction with the relation of \cite{2009MNRAS.400.2050G}; this corresponds to an extinction $A_V = 2.0^{+1.5}_{-1.0}$ mag (consistent with the nebular extinction within the errors), and an absolute brightness $M_V \approx -5.8$ mag, $M_I \approx -8.0$ mag (also consistent with a red supergiant).

In most of our sources, a model estimate of the de-absorbed X-ray luminosity is substantially affected by the large uncertainty on the absorbing column density (Table 6), and the degeneracy between absorption and normalization of the softest thermal plasma component. We used the {\sc xspec} model {\it cflux} to estimate the 90\% confidence limits of the absorbed and de-absorbed fluxes reported in Table 6.
As an order-of-magnitude estimate, the six thermal-plasma sources in our sample have unabsorbed luminosities $\sim$10$^{37}$ erg s$^{-1}$, with the possible exception of L14-63 (S1), which may be as luminous as $\approx$10$^{38}$ erg s$^{-1}$ after we account for its intrinsic absorption. The X-ray binary candidate L14-139 has an average X-ray luminosity $L_{\rm X} \approx 3 \times 10^{37}$ erg s$^{-1}$, consistent with the luminosity of the low/hard state.


\subsection{Radio Morphology and Fluxes}

By analogy with our discovery of MQ1 and of other microquasars, we are looking here for exceptionally luminous radio sources, and/or evidence of bipolar radio lobes. In this respect, the most interesting radio sources in our sample are S2 and S7. 


S2 is resolved into two sources of similar flux density, $S_{5.5\rm{GHz}} \approx 0.2$ mJy each, separated by $\approx$50pc. Both sources have a similar radio spectral index, $\alpha \approx -0.9 \pm 0.6$, consistent with optically thin synchrotron emission. The two radio sources roughly correspond to the two peaks of optical line emission identified by \cite{2012ApJS..203....8B} as the two separate SNR candidates B12-96 and B12-98. The hard, non-thermal X-ray source L14-139 is located approximately in between the two peaks of radio emission (Figure 2). The elongated region of optical line emission is also oriented along the same direction as the two radio sources, but extending well beyond them on either side. There are two possible interpretations for this configuration of radio, optical and X-ray emission: either a chance alignment of multiple SNRs (two of which are also radio bright), with an unrelated X-ray binary in between them; or a microquasar with a core corresponding to the point-like X-ray source, and radio/optical lobes produced by the interaction of the jet with the ISM.  



S7 is the brightest source in our sample, with a flux density $S_{5.5\rm{GHz}} \approx 1.0$ mJy, and one of the brightest compact radio sources in M\,83 (only slightly fainter than MQ1). Its flux measured from our 2011--2017 ATCA data is consistent with the flux measured by \cite{2006AJ....132..310M} from their 1998 VLA observations. Our ATCA observations reveal a spectral index of $\alpha = -0.35 \pm 0.28$: this is consistent  both with optically-thin synchrotron, and with free-free emission. We have already argued (Section 3.1), based on its optical line ratios, that S7 is associated with a photoionized H~{\footnotesize{II}} region, rather than a collisionally ionized bubble. Standard relations between Balmer emission and free-free radio emission in star-forming regions \citep{1986A&A...155..297C, 1992ARA&A..30..575C} predict that an H$\beta$ luminosity $\approx$1.5 $\times 10^{38}$ erg s$^{-1}$ should correspond to a 5.5-GHz flux density of $\approx$0.2 mJy at the distance of M\,83. In reality, the observed radio flux is 5 times higher. We do not have enough information to say whether S7 contains an additional radio synchrotron source (possibly a young SNR inside the H~{\footnotesize{II}} region), or whether the discrepancy is simply due to empirical scatter at the level of individual H~{\footnotesize{II}} regions. 


The radio source associated with S8 is another interesting case: the peak of the radio emission is displaced by $\approx0^{\prime\prime}.7$ south of the bright arc-like optical nebula (Figure 4), near the center point of the optical arc. The radio spectral index is consistent with flat ($\alpha \approx 0$),  although it could also be steep or inverted (Table 2). A possible interpretation of those two findings is that the optical arc is part of the expanding SNR shell, strongly interacting with the ISM only on its northern side because of a density gradient. Alternatively, the location and radio luminosity ($\sim$10$^{34}$ erg s$^{-1}$) of the radio source are consistent with a pulsar wind nebula, filling the central region of the SNR \citep{2006ARA&A..44...17G}. However, the associated X-ray source (L14-310) is centered somewhere between the optical and radio peaks (Figure 4), and is dominated by thermal plasma emission (Table 5).  Hence, the interpretation of this source remains uncertain.

Finally, we re-examined the radio emission from MQ1 (the brightest non-nuclear radio source in M\,83), taking advantage of the 2015 and 2017 ATCA observations that took place subsequent to the study of \cite{2014Sci...343.1330S}.  From our analysis of the combined ATCA dataset, we obtain integrated flux densities $S_{5.5\rm{GHz}} \approx 1.8$ mJy and $S_{9\rm{GHz}} \approx 1.0$ mJy, respectively. At the distance of M\,83, this corresponds to a radio luminosity of $\approx$2.5 $\times 10^{35}$ erg s$^{-1}$. This is much larger than the integrated 5.5-GHz radio luminosity of $\approx$5 $\times 10^{33}$ erg s$^{-1}$ for the most powerful microquasar bubble in the Milky Way, SS\,433/W50 \citep{1998AJ....116.1842D}; it is also higher than the most radio-luminous Milky Way SNR, Cassiopeia A, with a 5.5-GHz luminosity of $\approx$5.5 $\times 10^{34}$ erg s$^{-1}$ \citep{2018A&A...612A.110A}. 

According to \cite{2009ApJ...703..370C}, the radio luminosity function of SNRs is a power law and the number of SNRs is proportional to the star formation rate.  If that is the case, we expect the maximum radio luminosity of a SNR to be higher in M\,83 (with a star formation rate $\approx$3--5 $M_{\odot}$ yr$^{-1}$: \citealt{boissier05}) than in the Milky Way, with its more modest star formation rate ($\approx$1.5--2 $M_{\odot}$ yr$^{-1}$: \citealt{2015ApJ...806...96L}). As a result, we expect a few SNRs in M83 with a radio luminosity comparable to MQ1 even though none are observed in the Galaxy. The radio luminosity function of microquasar bubbles (powered by compact objects accreting in the super-critical regime) is not known yet, because of the small number of such sources identified so far; however, we do know of other bubbles with 5-GHz luminosities $\sim$10$^{35}$ erg s$^{-1}$ ({\it e.g.}, NGC\,7793-S26: \citealt{2010MNRAS.409..541S}; IC\,342 X-1: \citealt{2012ApJ...749...17C}). Although the radio luminosity and energy content of the most powerful radio SNRs and microquasar bubbles may be of the same order of magnitude, the two classes of sources differ in age and size: theory predicts \citep{2017MNRAS.464.2326S} that the most luminous radio SNRs have ages $\sim$10$^2$--10$^3$ yr and radii $\lesssim$a few pc ({\it e.g.}, the bright SNR in NGC\,4449: \citealt{2010MNRAS.409.1594B,2009AJ....137.3869C}), while accretion-powered bubbles can have ages $\gtrsim$10$^5$ yrs (timescale of super-critical mass transfer from the donor star) and sizes $\sim$100 pc \citep{2010Natur.466..209P,2006IAUS..230..293P,2002astro.ph..2488P}.


\section{Discussion } \label{sec:discussion}

We have seen that the majority of morphologically peculiar SNRs in our sample are consistent with  normal SNRs, based on their multiband properties ({\it e.g.}, optical line ratios and X-ray spectra). Instances where the optical line-emitting region appears linear or arc-like (specifically, S8 and S9) are best explained as enhanced interaction between an expanding SNR shell and the local ISM along one sector of the shell, perhaps because of density inhomogeneities. Another source (S6) has an intriguing linear structure in H$\alpha$ that does not appear to be part of a larger spherical shell; however, its faintness and lack of an X-ray counterpart prevent detailed conclusions on its SNR versus microquasar nature. 

The two sources that most stand out in our sample are S2 and S7. S2 is either a series of unrelated SNRs, aligned by chance also with an X-ray binary; or it is a single physical object, a jet-powered bubble, $\approx$200 pc in length, with an optically emitting bow shock at its western end and a series of internal shocks closer to the core. In the latter scenario, the non-thermal X-ray source is the microquasar core (origin of the jet). A microquasar jet with similar morphology was recently discovered in NGC\,300 \citep{2019MNRAS.482.2389U,2019MNRAS.485.3476M}. If this scenario is correct, the pair of optical SNR candidates B12-96 and B12-98 and the pair of synchrotron radio sources at the same location correspond to the optical/radio lobes for the current phase of jet activity, while the fainter H$\alpha$ emission beyond those lobes corresponds to earlier phases of activity.


Standard bubble theory \citep{1977ApJ...218..377W} shows that the total radiative luminosity $L_{\rm rad}$ of shock-heated gas is $L_{\rm rad} \approx (27/77) P_{\rm kin}$, where $P_{\rm kin}$ is the kinetic power of the wind or jet that is inflating the bubble. (The rest of the kinetic power is spent for the bulk motion of the expanding swept-up shell, and for the work done to expand the bubble against the external ISM pressure.) In turn, the relative contribution to the radiative luminosity from the various IR/optical/UV lines depends on the undisturbed ISM density, the shock velocity, the metallicity, and the magnetic energy density, and can be calculated with codes such as {\sc mappings} \citep{1995ApJ...455..468D,2008ApJS..178...20A}. In particular, the fractional contribution of the H$\beta$ line emission (total emission from shock and precursor) does not depend too strongly on the shock velocity, and can be approximated as $L_{{\rm H}\beta} \approx 2.5 \times 10^{-3} P_{\rm{kin}}$ \citep{1995ApJ...455..468D,2010Natur.466..209P,2014Sci...343.1330S}, within a factor of two, over the range of typical microquasar shock velocities ($v_{\rm s} \approx 100$--300 km s$^{-1}$). Thus, assuming that the line emission from S2 is due to jet-driven shocks, and taking $L_{{\rm H}\alpha} \approx 3.0 L_{{\rm H}\beta}$\footnote{The Balmer decrement for radiative shocks is slightly steeper than the canonical value of 2.86 suitable to photo-ionized nebulae: for solar metallicity and a shock velocity of 150 km s$^{-1}$, the H$\alpha$/H$\beta$ ratio is 3.06, while for a shock velocity of 500 km s$^{-1}$, the ratio is 2.92: \citep{2008ApJS..178...20A}}, we estimate $P_{\rm kin} \approx 2.6 \times 10^{40}$ erg s$^{-1}$ from the total emission of the 200-pc nebula, or, perhaps more appropriately, $P_{\rm kin} \approx 1.5 \times 10^{40}$ erg s$^{-1}$ if we consider only the inner pair of shocked regions (B12-96 and B12-98), interpreted as the currently active jet lobes (by analogy with \citealt{2019MNRAS.485.3476M} and \citealt{2019MNRAS.482.2389U}). This is similar to the kinetic power estimated for MQ1 from its line emission \citep{2014Sci...343.1330S}.

The other outstanding source in our sample is S7. This is the only nebula in our list that is clearly dominated by UV photo-ionization, based on its line ratios. We showed that a young star cluster with a mass $\sim$10$^4 M_{\odot}$ and an age $\lesssim$3 Myr (thus, still containing dozens of O stars) is consistent both with the observed optical continuum and with the ionizing photon flux required to explain the huge H$\alpha$ emission ($L_{\rm{H}\alpha} \approx 4 \times 10^{38}$ erg s$^{-1}$). [Fe {\footnotesize{II}}] emission is extended ($\approx$30 pc across) and relatively weak compared with H$\alpha$ (as expected in a photoionized region), but still quite strong in absolute terms:  $L_{\rm{[Fe~II]}} \approx 5 \times 10^{36}$ erg s$^{-1}$. Possible non-SNR origins for the [Fe {\footnotesize{II}}] emission are protostellar and stellar outflows \citep{2014ApJS..214...11S,2016MNRAS.463.4344R}, and the shock front where the H {\footnotesize{II}} region advances into the surrounding neutral ambient medium \citep{1977ApJ...214..725E,spitzer1978,2018MNRAS.479.2016W}. However, it is also possible that S7 contains a buried SNR, whose contribution to the Balmer emission may be negligible (compared with the photo-ionized contribution) but which may explain the [Fe {\footnotesize{II}}] emission. Another peculiar property of S7 is its strong radio emission, with a flat spectral index but a luminosity about 5 times higher than expected for free-free radio emission in an H {\footnotesize{II}} region. That may be another clue about the presence of a buried SNR inside the H {\footnotesize{II}} region.

For some of the other SNR candidates in our list, the simplest explanation for their irregular morphology is an asymmetry of the surrounding CSM/ISM. For example, in S4, S8 and S9, the SN shock wave must be expanding into denser ISM towards the northern part of the remnant. Conversely, in S3, the ISM is likely to have a density gradient towards the southwest. S3 is also notable in the {\it HST} images (Figure 2) for its high ionization, based on the observed strong [O {\footnotesize{III}}] emission: the unabsorbed $L_{\rm{[O III]}} \approx 3 L_{\rm{H}\alpha}$ (Table 4) likely indicates a high shock velocity for this object, especially for the northern extension{\footnote{The alternative of an ejecta-dominated SNR similar to Cas A (or  SN1957D in M83) can be excluded by the fact that the velocity widths of $\approx$400 km s$^{-1}$ are much smaller than for ejecta-dominated SNRs, and are the same for all the lines, including H$\alpha$ and H$\beta$.}}  
Finally, S6 has a collisionally ionized linear structure also reminiscent of a microquasar jet, but no X-ray detection and only a marginal radio detection at 5.5 GHz. If it is a (currently inactive) microquasar, applying again the line luminosity scaling to its H$\alpha$ and [Fe {\footnotesize{II}}]$\lambda$1.64$\mu$m emission \citep{2010Natur.466..209P,2014Sci...343.1330S}, we obtain a kinetic power $\approx$10$^{39}$ erg s$^{-1}$, but there is no evidence ruling out an SNR interpretation.

\section{Conclusions } \label{sec:conclusions}


In this study, we have selected a sample of nine irregularly shaped SNR candidates in M\,83---a purely empirical assessment based on their appearance in {\it HST}/WFC3 images. We have analyzed their multiband properties, combining {\it Chandra}, {\it HST}, Gemini and ATCA data, with the objective of determining the nature of those sources and possibly discovering a new microquasar among them (by analogy with our earlier discovery of MQ1). We found that at least six of them (S1, S3, S4, S5, S8, S9) are indeed best interpreted as SNRs, based on their optical and infrared line ratios and their soft, thermal-plasma X-ray spectrum ($kT \lesssim 1.0$ keV). For one source (S6) we do not have enough observational data to explain the reason of its linear morphology. One optical nebula (S7) is dominated by photoionization. The origin of the ionizing photons appears to be a cluster of O stars. It is plausible that star formation in that region is so recent that it has not produced SNe yet; however, an intriguing property of S7 is its strong radio luminosity and the presence of [Fe {\footnotesize{II}}] emission, which may suggest the presence of a hidden SNR. We leave further analysis of the radio source populations associated with SNRs and with H{\footnotesize{~II}} regions to a future study (T.~D.~Russell et al., in prep.) 

The source labelled S2 is the most complex and perhaps most interesting. It is the only source in our sample for which we propose a possible microquasar jet interpretation. With the data currently available, we cannot distinguish between the microquasar and the multiple SNR scenario. In the microquasar scenario, S2 is a single physical structure powered by the jet from an accreting compact object (the {\it Chandra} source L14-139), which produces optical and radio lobes and an elongated tail of shock-ionized gas. In the multiple SNR scenario, the linear structure is caused by a chance alignment of several SNRs, and the core is an unrelated X-ray binary, also a aligned by chance. 

If S2 is a microquasar, its kinetic power is $\sim$10$^{40}$ erg s$^{-1}$, in the same class as the most powerful super-critical jet sources found so far in the local universe. The X-ray luminosity of the candidate core is currently either low ($\sim$10$^{37}$ erg s$^{-1}$) or obscured by Compton-thick material, so that we  see only scattered emission (as is the case for SS\,433). Such sources are rare (even rarer than the radiatively bright ultraluminous X-ray sources) but are a key test of super-critical accretion models. To make further progress, we need to determine whether there is a systematic positive and negative velocity shift in the optical emission lines on either side of the candidate core (signature of a large-scale outflow). We have obtained a set of spectra with the Multi Unit Spectroscopic Explorer (MUSE) on the Very Large Telescope: we will present the results in a forthcoming work.

\acknowledgments
We thank Anna McLeod, Christian Motch, Manfred Pakull, and Ryan Urquhart for helpful discussions. We also thank James Miller-Jones for his help with the planning and analysis of the radio observations. The final version of this paper was much improved thanks to competent comments from the referee. This paper is dedicated to the memory of Michael A. Dopita (1946--2018), who pioneered modelling of optical emission from shock-heated plasmas, who started the morphological classification of SNRs in M\,83 from the initial  {\it HST} survey, and who will be much missed by all of us. RS acknowledges support and hospitality from the Curtin Institute of Radio Astronomy (Perth, Australia) during the early stages of this work.  WPB acknowledges STScI grant HST-GO-14211.002-A for partial support during this project.  PFW acknowledges  support from NASA  grant HST-GO-14211.003 and from NSF grant AST-1714281.

\label{lastpage}


\begin{table*}
	\centering
	\scriptsize
	\caption{Nine selected SNR candidates with peculiar morphology, and comparison jet source MQ1}	
        \label{tab:source_id}
        \vspace{0.3cm}
	\begin{tabular}{lccccc} 
		\hline\\
		Source ID & RA & Dec & \multicolumn{3}{c}{Cross-ID$^{a}$}\\[5pt]
		    &    &    &    X-ray & Optical & Radio \\[5pt]
		\hline\\
		S1 & 13:36:50.85 & $-$29:52:39.6 
		                & L14-63 & W19-49, B12-45 & L14-A12, M06-03\\[5pt]
		S2 & 13:36:55.56 & $-$29:53:03.5 & L14-139 
		         & W19-\{112+114\}, B12-\{96+98\}, BL04-24 
		         & L14-A39, M06-22\\[5pt]
		S3 & 13:36:59.33 & $-$29:55:08.9 &  L14-183 & W19-145, B12-122 & \\[5pt]
		S4 & 13:36:59.50 & $-$29:52:03.7 & L14-186 & W19-150, B12-124, D10-04 
		       &  L14-A57\\[5pt]
		MQ1 & 13:37:01.12 & $-$29:51:52.2 & L14-237
		        & W19-186, S14-MQ1,   D10-N16 &     L14-A62 \\[5pt]
		S5 & 13:37:01.73 & $-$29:51:13.4 & L14-256  
		   &  W19-195, B12-143, D10-12, BL04-37 \\[5pt]	
		S6 & 13:37:02.12 & $-$29:51:58.8  & 
		   & W19-199, B12-146, D10-14, BL04-39 & \\[5pt]
		S7 & 13:37:03.44 & $-$29:54:02.5 & 
		     & W19-207$^a$, B14-45, SL11-NGC\,5236-2-530    &   L14-A72, M06-38 \\[5pt]
		S8 & 13:37:06.03 & $-$29:55:14.3 &  L14-310 
		     & W19-231, B12-169, BL04-46  & \\[5pt]
		S9 & 13:37:11.87 & $-$29:52:15.6 & L14-358 
		      &  W19-286, B12-209 & \\[5pt]
		\hline\\[3pt]
	\end{tabular}\\
			$^a$W19 = \cite{williams19};  B14 = \cite{2014ApJ...788...55B}; L14 = \cite{2014ApJS..212...21L}; S14 = \cite{2014Sci...343.1330S}; B12 = \cite{2012ApJS..203....8B}; SL11 = \cite{2011A&A...529A..25S}; D10 = \cite{2010ApJ...710..964D}; M06 = \cite{2006AJ....132..310M}; BL04 = \cite{2004ApJS..155..101B}.
			\vspace{0.5cm}
\end{table*}

\begin{table*}
	\centering
	\caption{Observed fluxes/count rates in various bands: Data from {\it Chandra}, {\it HST}, and ATCA}	
        \label{tab:hst_log_table}
        \vspace{0.3cm}
	\scriptsize
	\begin{tabular}{lccccccc} 
		\hline\\
		ID  
                & Ct Rate (0.35--8 keV) & $F_{\rm{[O~III]}}$ & $F_{\rm{H}\alpha +[N~II]}$ & $F_{\rm{[S~II]}}$ & $F_{\rm{[Fe~II]}}$ & 
                $S_{\rm 9GHz}$ & $S_{\rm 5.5 GHz}$  \\[5pt]
		 &  ($10^{-3}$ ct s$^{-1}$) & ($10^{-15}$ CGS) & ($10^{-15}$ CGS)  &
		 ($10^{-15}$ CGS)  & ($10^{-15}$ CGS)  &
		 ($\mu$Jy\,beam$^{-1}$) & ($\mu$Jy\,beam$^{-1}$)  \\[5pt]
		\hline\\
		S1 & $0.154\pm0.021$ & $0.83\pm0.08$ & $3.8 \pm 0.5$ & $1.3\pm 0.1$ & $0.50\pm 0.05$ & $140 \pm 18$  & $200 \pm 15 $ \\[5pt]
		S2 & $0.455\pm0.028$ & $2.6 \pm 0.9$ & $23.6 \pm 2.0$& $7.7 \pm 1.1$ & --- & $71 \pm 12$ (East) & $120 \pm 20$ (East)  \\[5pt]
		 & &  & &  & & $82 \pm 12$ (West) & $130 \pm 20$  (West) \\[5pt]
		S3 & $0.186\pm0.020$  & $2.3 \pm 0.2$ & $3.1 \pm 0.3$ & $0.86\pm0.08$ &  $0.29\pm 0.03$ & $60 \pm 10$ & $59 \pm 10$ \\[5pt]
		S4 & $0.327\pm0.026$ & $1.2\pm 0.1$ & $5.1\pm0.5$ & $1.8\pm0.2$ & $0.40\pm0.04$ & $190 \pm 20$  & $280 \pm 20 $\\[5pt]
        MQ1 & $1.823\pm0.052$ & $0.35 \pm 0.04$ & $5.2 \pm 0.6$ & $1.7 \pm 0.3$ & $4.4 \pm 0.4$ & $1000 \pm 200$ & $1800 \pm 200$ \\[5pt]
		S5 & $0.234\pm0.019$ & $2.6 \pm 1.0$ & $8.0 \pm 1.3$ & $2.9 \pm 0.7$ & $0.81 \pm 0.07$  & $<80$ & $82 \pm 20$  \\[5pt]
		S6 & $<0.02$& $0.43 \pm 0.28$& $4.3 \pm 0.8$ & $2.4 \pm 0.5$ & $0.4 \pm 0.1$ & $<90$ & $110 \pm 20$   \\[5pt]
		S7 & $<0.02$& $2.4 \pm 0.9$ & $50 \pm 4$ & $3.7 \pm 1.2$ & $1.2 \pm 0.2$ & $850 \pm 15$ & $1000 \pm 15$  \\[5pt]
		S8 & $0.576\pm0.030$ & $1.2 \pm 0.7$ & $5.4 \pm 1.2$ & $1.7 \pm 0.7$& $0.78 \pm 0.08$ & $52 \pm 15$ & $54 \pm 10$  \\[5pt]
		S9 & $0.416\pm0.025$ & $1.7\pm0.2$ & $6.7\pm0.6$ & $1.4\pm0.2$ & $0.37\pm0.04$ & $<33$ & $<48$ \\[5pt]
		
		\hline
	\end{tabular}
\end{table*}


\begin{table*}
	\centering
	\caption{Line fluxes, line ratios, densities and extinction from the Gemini spectra.}	
        \label{tab:gemini_fluxes}
        \vspace{0.3cm}
        \scriptsize
	\begin{tabular}{lccccccccc} 
		\hline\\
		ID  & 
		$F_{\rm{[O~III]}}$ &
		$F_{\rm{H}\alpha}$ &
		$F_{\rm{H}\alpha}/F_{\rm{H}\beta}$ &
		$F_{\rm{[S II]}}/F_{\rm{H}\alpha}$ &
		$F_{\rm{H}\alpha}/F_{\rm{H}\alpha +[N~II]}$ &
		$F_{6716}/F_{6731}$ & $n_e$ &
		$A_V$ &
		Ionization \\[5pt]
		 & ($10^{-15}$ CGS) & ($10^{-15}$ CGS) & & & & & (cm$^{-3}$) & (mag) &  \\[5pt]
		\hline\\
		S1 & 1.0 & 4.2 & 4.69 & 0.48 & 0.53 & 1.22 & 180 & 1.55 & S\\[5pt]
		S2 & 1.6 & 5.6 & 6.67 & 0.81 & 0.44 & 1.33 & 84 & 2.66 & S\\[5pt]
		S3 & 2.2 & 1.4 & 4.76 & 0.69 & 0.46 & 1.17 & 240 & 1.60 & S\\[5pt]
		S4 & 0.66 & 0.73 & 5.45 & 1.62 & 0.23 & 1.02 &  460 & 2.03 & S\\[5pt]
		S6 & 0.31 & 1.2 & 4.35 & 1.47 & 0.33 & 1.35 & 70 & 1.31 & S\\[5pt]
		S7 & 3.8 & 72.1 & 6.00 & 0.19 & 0.59 & 1.15 & 260 & 2.32 & P\\[5pt]
		S8 & 0.95 & 1.65 & 4.92 & 0.99 & 0.32 & 0.89 & 770 &  1.70 & S\\[5pt]
		S9 & 1.8 & 3.0 & 3.70 & 0.58 & 0.49 & 1.06 & 390 & 0.81 & S\\[5pt]
		\hline
	\end{tabular}\\[5pt]
	{\it Notes}: S5 and MQ1 were not observed with Gemini.
	\vspace{0.5cm}
\end{table*}

\begin{table}
	\centering
	\caption{Extinction-corrected line luminosities from the {\it HST} data.}	
        \label{tab:hst_luminosity}
        \vspace{0.3cm}
        \scriptsize
	\begin{tabular}{lccc} 
		\hline\\
		ID  
                &  $L_{\rm{[O~III]}}$   & $L_{\rm{H}\alpha}$  & $L_{\rm{[Fe~II]}}$  \\[5pt]
		 & ($10^{37}$ erg s$^{-1}$) & ($10^{37}$ erg s$^{-1}$) & ($10^{37}$ erg s$^{-1}$) \\[5pt]
		\hline\\
		S1 & $1.0\pm0.1$ & $1.6\pm0.2$ & $0.12 \pm0.01$\\[5pt]
		S2  & $10.3 \pm 3.5$ & $19.6 \pm 1.7$ & -- \\[5pt]
		S3 & $3.0\pm0.3$ & $1.2\pm0.1$ & $0.10 \pm 0.01$\\[5pt]
		S4 & $2.5\pm0.3$ & $1.4\pm0.2$ & $0.14\pm0.02$\\[5pt]
		S5 & $3.8 \pm 1.5$ & $2.9 \pm 0.5$ & $0.27 \pm 0.03$  \\[5pt]	
		S6 & $0.4 \pm 0.3$  & $1.0 \pm 0.2$ & $0.13 \pm 0.03$\\[5pt]
		S7 & $6.7 \pm 2.5$ & $43.1 \pm 3.4$ & $0.45 \pm 0.07$   \\[5pt]
		S8 & $1.8 \pm 1.0$ & $1.6 \pm 0.3$ & $0.26 \pm 0.03$    \\[5pt]
		S9 & $1.0\pm0.1$ & $1.5\pm0.2$ & $0.11 \pm 0.01$\\[5pt]
		MQ1 & $5.0 \pm 0.5$ & $10.0 \pm 1.2$ & $2.1 \pm 0.2$\\[5pt]
		\hline
	\end{tabular}
	\vspace{0.5cm}
\end{table}

\begin{table*}
	\centering
	\caption{Observed flux from the microquasar candidate S2 from the individual {\it Chandra} observations}	
        \label{tab:S2_flux_table}
        \vspace{0.3cm}
        \scriptsize
	\begin{tabular}{lcccc} 
		\hline\\
		ObsID & MJD(start)  & Date & $F_{0.5-7}$ & Model $F_{0.5-7}$ \\[5pt]
	     & &  & ($10^{-15}$ erg cm$^{-2}$ s$^{-1}$)  & ($10^{-15}$ erg cm$^{-2}$ s$^{-1}$) \\[5pt]
		\hline\\
		793 & 51663.58 &  2000-04-29  & $7.2^{+1.8}_{-1.8}$  
		    & $8.8^{+2.2}_{-2.2}$\\[5pt]
		2064 & 52157.00 &  2001-09-04 & $14.5^{+10.4}_{-7.6}$  
		    & $7.7^{+5.5}_{-4.1}$\\[5pt]
		12995 & 55553.43 &  2010-12-23  & $3.0^{+1.1}_{-1.0}$ 
		    & $5.0^{+1.9}_{-1.6}$  \\[5pt]
		13202 & 55555.69 &  2010-12-25  & $6.7^{+1.5}_{-1.6}$ 
		    & $6.5^{+1.5}_{-1.5}$ \\[5pt]
		12993 & 55635.50 &  2011-03-11 & $3.8^{+1.5}_{-1.1}$
		    & $6.1^{+2.3}_{-1.8}$  \\[5pt]
		13241 & 55638.90 &  2011-03-18  & $5.7^{+1.3}_{-1.4}$
		    & $7.3^{+1.8}_{-1.7}$ \\[5pt]
		12994 & 55643.15 &  2011-03-23  & $4.1^{+0.8}_{-0.8}$
		    &  $6.4^{+1.3}_{-1.2}$ \\[5pt]
		12996 & 55649.67 &  2011-03-29  & $4.8^{+2.5}_{-1.9}$
		    & $3.7^{+1.8}_{-1.5}$ \\[5pt]
		13248 & 55654.30 &  2011-04-03 & $3.9^{+1.7}_{-1.4}$ 
		    & $4.4^{+1.9}_{-1.5}$\\[5pt]
		14332 & 55802.77 &  2011-08-29 & $1.7^{+1.1}_{-0.9}$ 
		    & $2.2^{+1.4}_{-1.1}$\\[5pt]
		12992 & 55808.25 &  2011-09-04 & $3.6^{+1.4}_{-1.2}$ 
		    & $3.6^{+1.4}_{-1.1}$\\[5pt]
		14342 & 55923.41 &  2011-12-28 & $6.0^{+1.9}_{-1.6}$ 
		    & $6.0^{+1.9}_{-1.6}$\\[5pt]
		\hline
	\end{tabular}\\[5pt]
	{\it Notes}: model-independent and model-dependent fluxes calculated with the {\sc ciao} task {\it srcflux}. The fluxes are not corrected for absorption. 
	\vspace{0.5cm}
\end{table*}

\begin{table*}
\tiny
\begin{center}
	\caption{Best-fitting model parameters and luminosities of the X-ray counterparts}	
        \label{tab:chandra_spectra}
        \vspace{0.3cm}
	\begin{tabular}{lccccccccccc} 
		\hline\\
		ID  & $(N_{\rm H})^a$ 
                &  $kT_1$   & $(N_{1})^b$  & $kT_2$  & $(N_2)^b$ & $\Gamma$ & $(N_{\rm po})^c$
                    & $\chi^2_{\nu}$  &  $F_{0.3-10}$ & $L_{0.3-10}$  \\[5pt]
		 & ($10^{22}$ cm$^{-2}$) & (keV) & ($10^{-6}$ cm$^{-5}$) & (keV) & ($10^{-6}$ cm$^{-5}$) & & ($10^{-7}$) & & ($10^{-15}$ CGS)  &  ($10^{37}$ CGS)  \\[5pt]
		\hline\\
		L14-63 (S1) & $0.71^{+0.22}_{-0.40}$ & $0.19^{+0.08}_{-0.04}$ & $21.0^{+86.3}_{-18.6}$ & - &- & - &- & 0.41 (5.7/14)  & $1.3^{+0.2}_{-0.2}$  &  $10.8^{+41.6}_{-9.4}$ \\[5pt]
		L14-139 (S2) & $0.39^{+0.33}_{-0.23}$ & - & - & $0.52^{+0.12}_{-0.21}$ & $1.0^{+3.7}_{-0.6}$    & $1.27^{+0.31}_{-0.30}$  &  $6.3^{+2.5}_{-1.9}$ &  0.91 (21.8/24) & $6.7^{+1.3}_{-1.1}$ & $2.5^{+2.2}_{-0.6}$  \\[5pt]
		L14-183 (S3) & $0.13^{+0.36}_{-0.13}$ & $<0.21$ & $8.3^{+87.0}_{-8.2}$ & $0.58^{+0.09}_{-0.11}$  &  $0.47^{+0.54}_{-0.18}$ & - & -& 1.18 (8.28/7)  & $1.1^{+0.7}_{-0.3}$  & $0.9^{+5.8}_{-0.6}$  \\[5pt]
		L14-186 (S4) & $0.18^{+0.55}_{-0.18}$ & $<0.20$ & $25^{+660}_{-24}$ & $0.75^{+0.15}_{-0.16}$  &  $0.93^{+1.44}_{-0.36}$ & -  & -  &  1.07 (22.52/21) & $2.0^{+0.8}_{-0.4}$ & $2.3^{+43.5}_{-1.8}$  \\[5pt]
		L14-256 (S5) & $0.07^{+0.56}_{-0.07}$ & $0.27^{+0.07}_{-0.08}$ & $0.58^{+11.9}_{-0.23}$  &  $0.99^{+0.21}_{-0.21}$ & $0.36^{+0.28}_{-0.12}$  &  - & -  & 0.82 (9.0/11) & $1.4^{+0.2}_{-0.2}$ & $0.52^{+6.0}_{-0.15}$   \\[5pt]
		L14-310 (S8) & $<0.31$ & $0.26^{+0.05}_{-0.08}$ & $0.89^{+0.25}_{-0.28}$ &  $0.80^{+0.13}_{-0.12}$ &    $0.87^{+0.24}_{-0.22}$ & -  &  - & 0.88 (18.48/21) & $3.5^{+0.3}_{-0.3}$ & $1.0^{+2.6}_{-0.1}$  \\[5pt]
		L14-358 (S9) & $0.05^{+0.11}_{-0.05}$ & $<0.12$ & $11.9^{+4.3}_{-10.3}$ & $0.49^{+0.07}_{-0.05}$  &    $0.85^{+0.27}_{-0.18}$ &   -  &  -  & 0.91 (14.61/16) & $2.7^{+0.4}_{-0.4}$  & $1.4^{+1.3}_{-0.6}$  \\[5pt]
		L14-237 (MQ1) & \multicolumn{10}{c}{See detailed spectral modelling in \cite{2014Sci...343.1330S}}    \\[5pt]
		\hline
	\end{tabular}\\[5pt]
	\end{center}
	    $^a$: intrinsic absorption column density. In addition, a fixed Galactic foreground column density $N_{\rm H} = 4 \times 10^{20}$ cm$^{-2}$ was included in every model.
	
		$^b$: standard {\it mekal} normalization $N = 10^{-14}/(4 \pi d^2) \, \int n_e n_{\rm H} dV$,  where $n_e$ and $n_H$ are the electron and hydrogen densities in the emitting plasma (units of cm$^{-3}$), and $d = 4.61$ Mpc $= 1.42 \times 10^{25}$ cm.
		
		$^c$: standard power-law normalization in units of photons (keV)$^{-1}$ cm$^{-2}$ s$^{-1}$ at 1 keV.
	\vspace{0.5cm}
\end{table*}

\newpage

\begin{figure}
    \centering
    \includegraphics[width=0.97\textwidth]{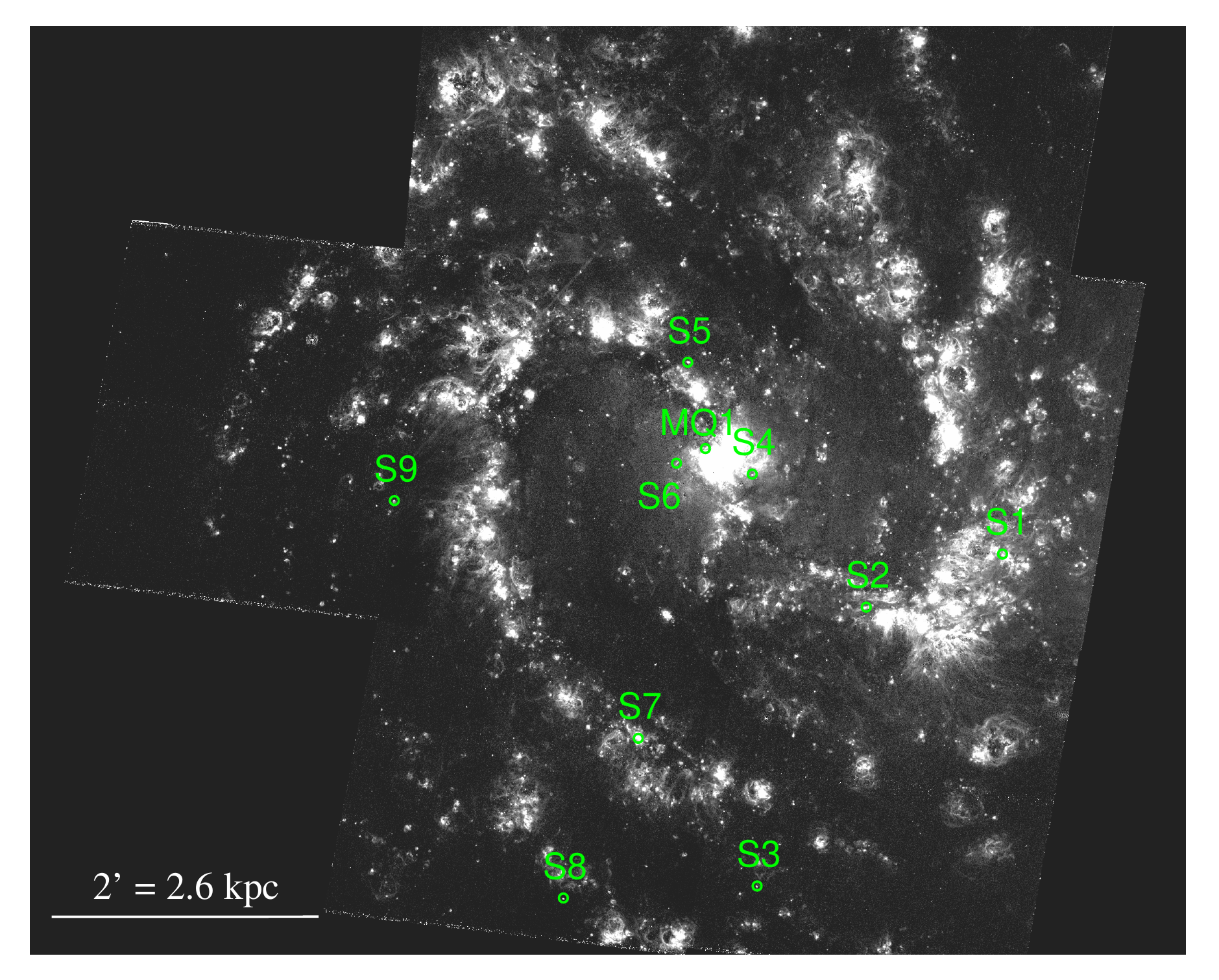}
\caption{Finding chart for the 9 candidate SNRs discussed in this paper (S1 through S9); the location of MQ1 is also plotted. The greyscale image is a continuum-subtracted {\it HST}/WFC3 image in the F657N filter; see \cite{2014ApJ...788...55B} for details. Note that several of the objects are clustered on the outskirts of the bright star-forming nuclear region.}
    \vspace{0.5cm}
\end{figure}

\begin{figure*}
    \centering
    \vspace{-0.2cm}
    \includegraphics[width=0.495\textwidth]{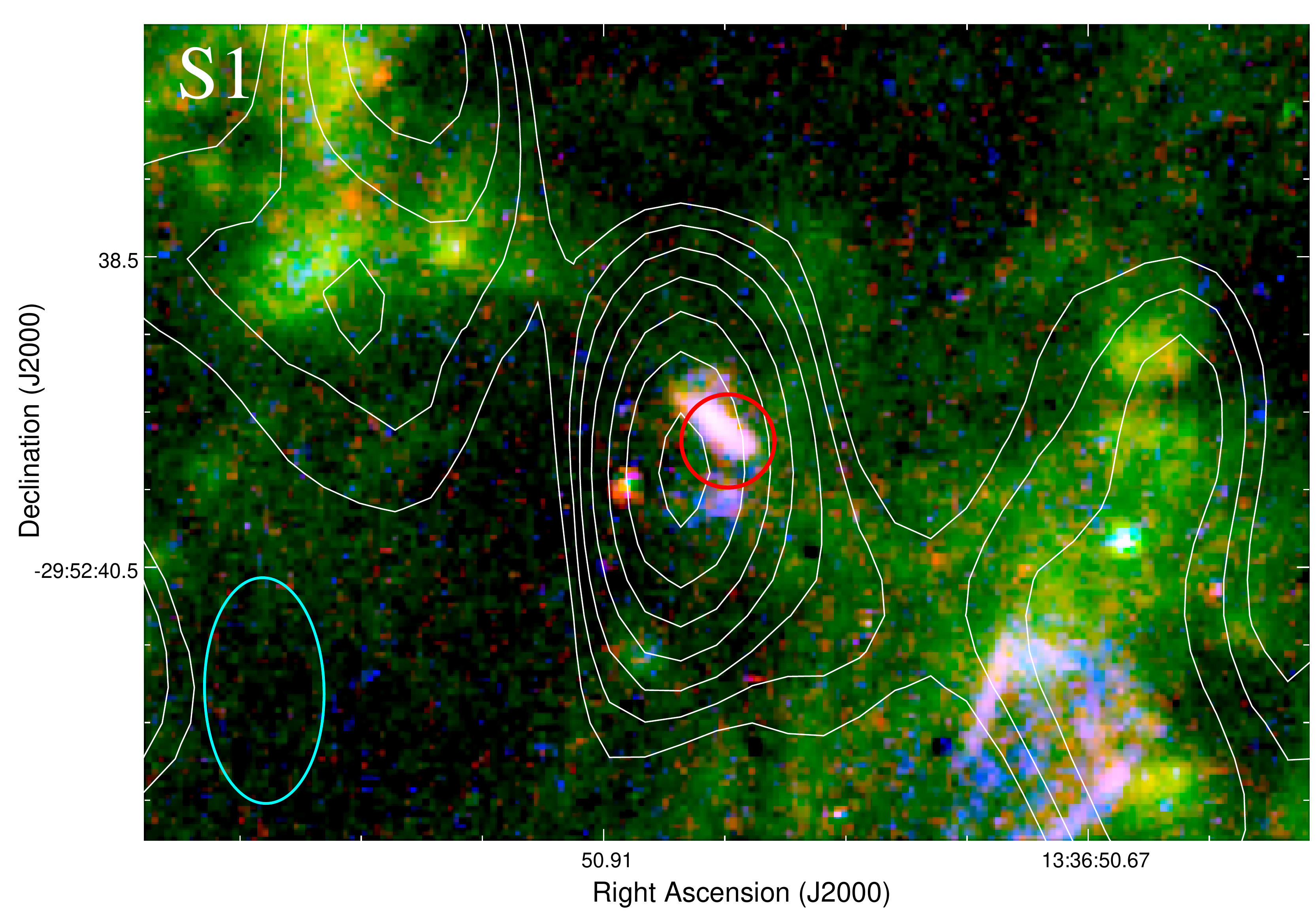}
    \includegraphics[width=0.495\textwidth]{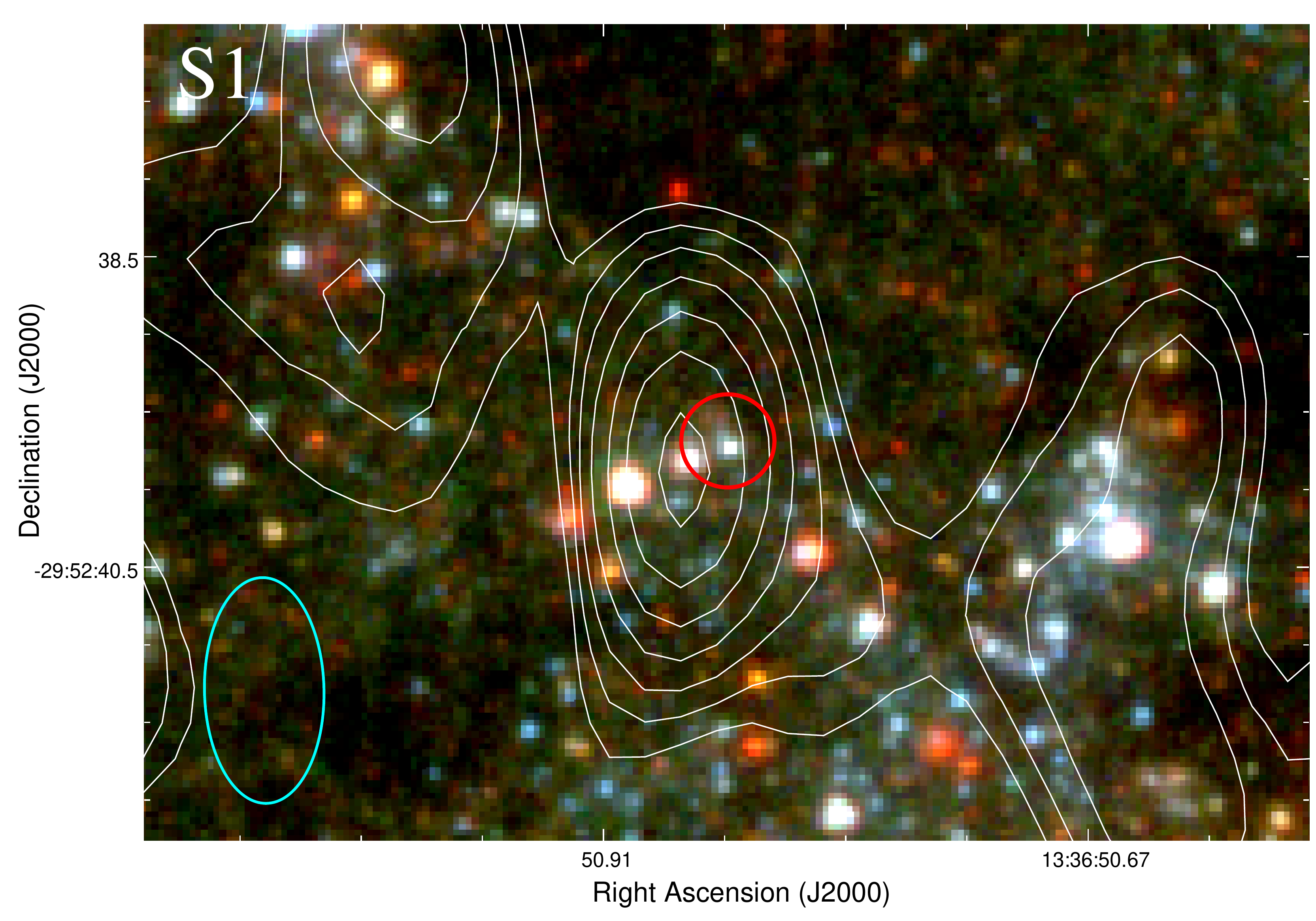}
   \\[-3pt]
    \includegraphics[width=0.495\textwidth]{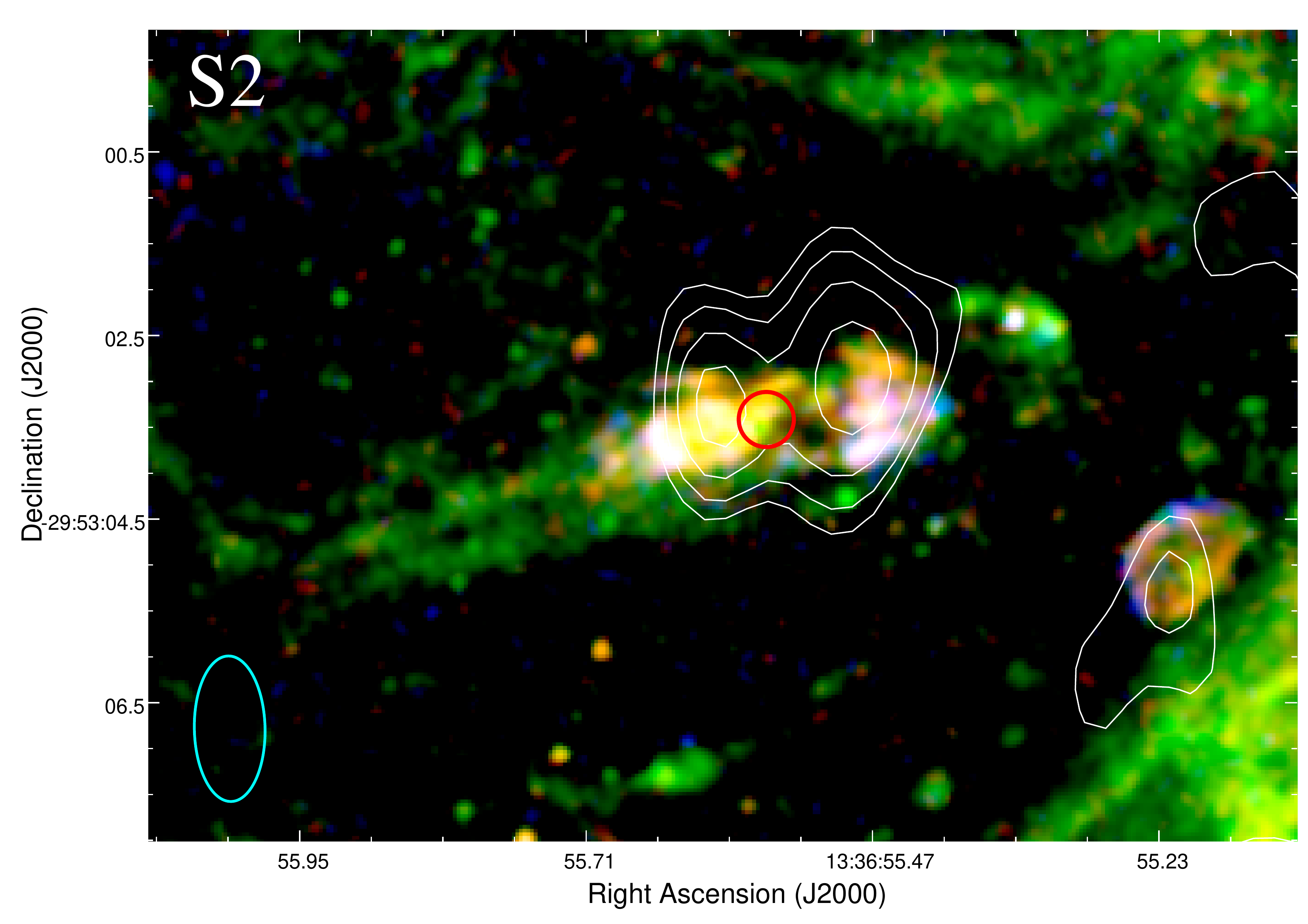}
    \includegraphics[width=0.495\textwidth]{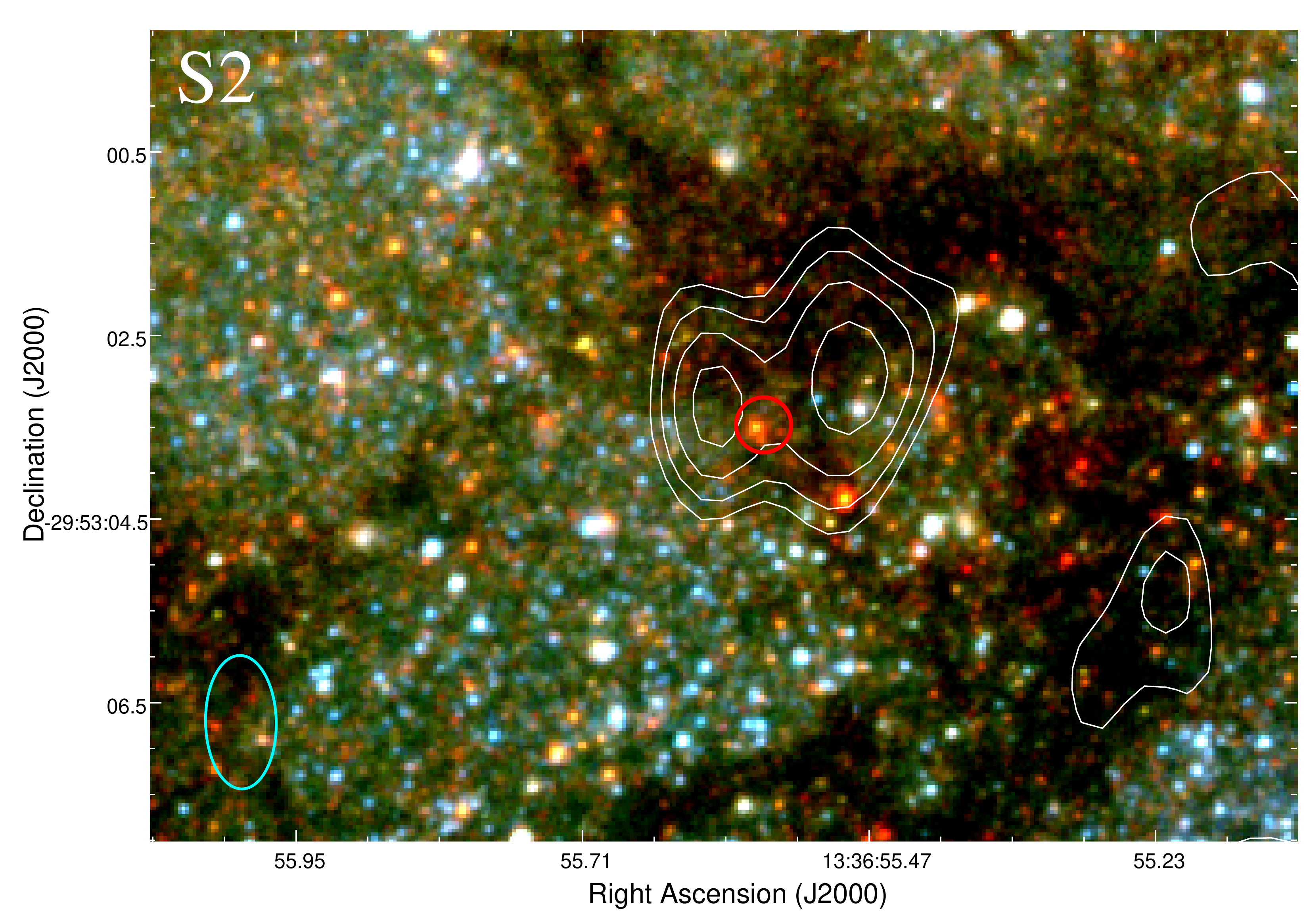}
    \\[-3pt]
    \includegraphics[width=0.495\textwidth]{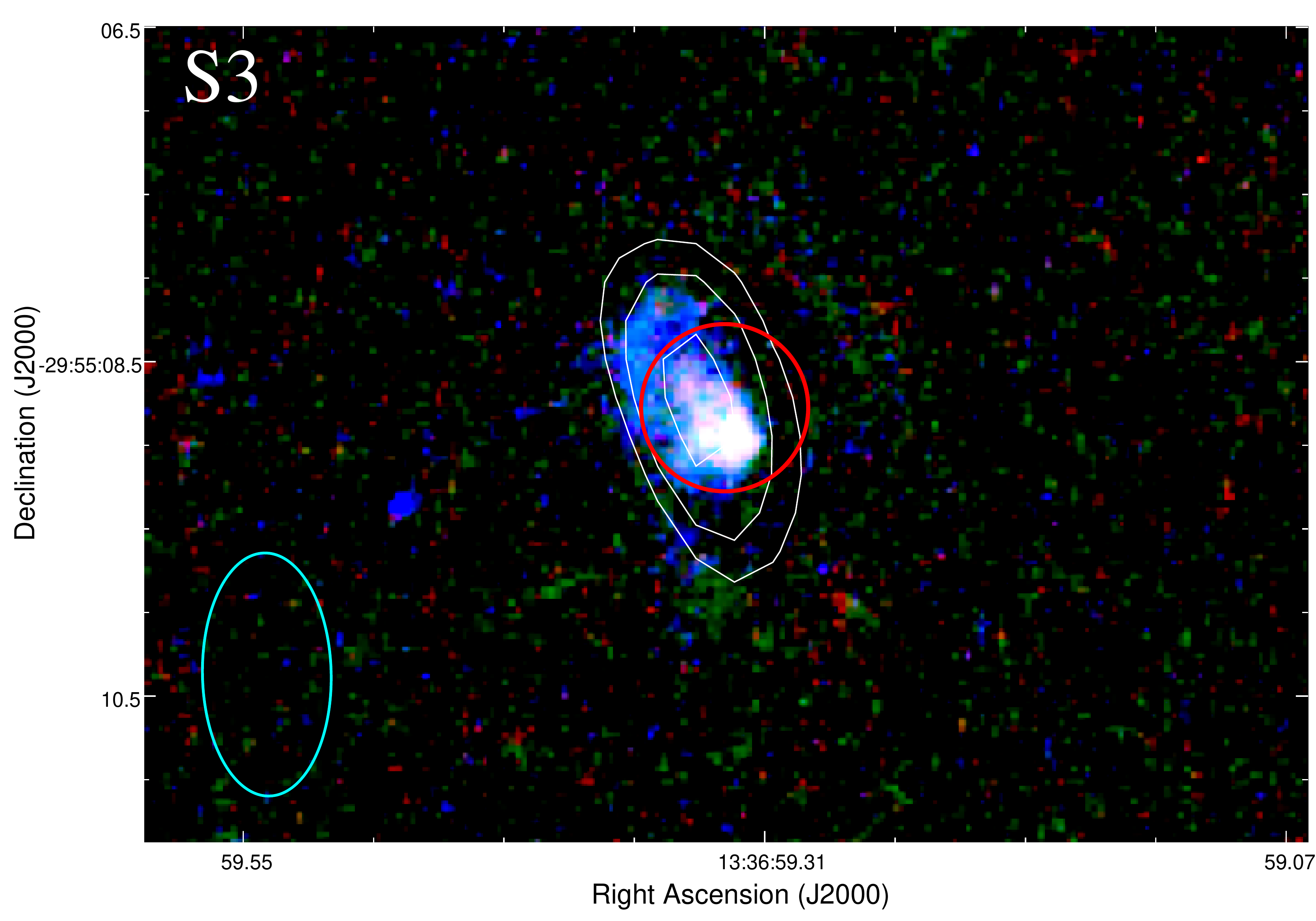}
    \includegraphics[width=0.495\textwidth]{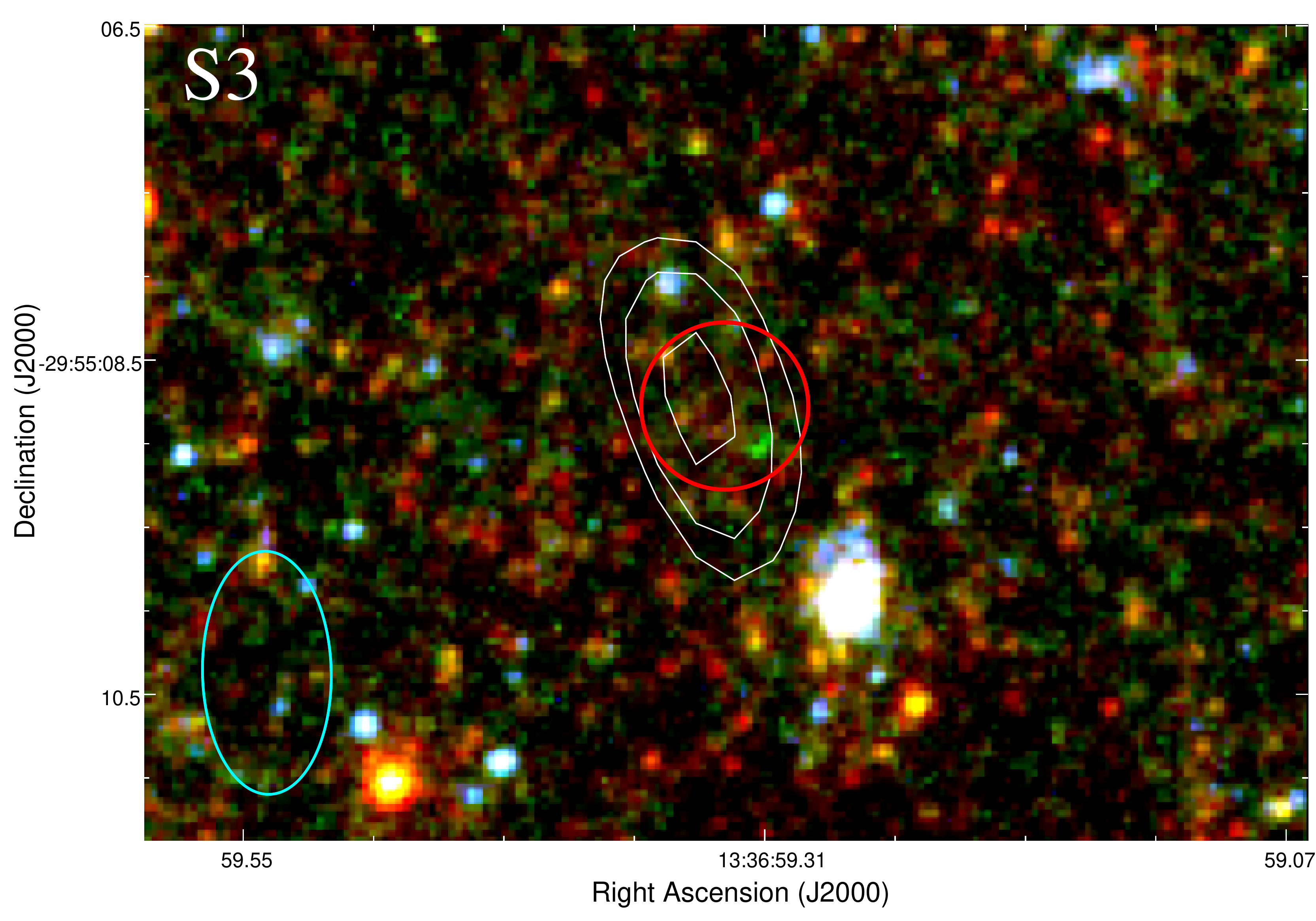}
\vspace{-0.7cm}
\caption{
Left panels: continuum-subtracted narrow-band {\it HST}/WFC3 images for candidate SNRs S1, S2, S3 (see Table 1 for identification), where red is F673N, green is F657N, and blue is F502N. Right panels: the same regions in WFC3 continuum bands, where red is F814W, green is F547M, and blue is F438W. In both sets of panels, red circles represent the centroid of the corresponding {\it Chandra} sources. The radius of the {\it Chandra} error circle is approximately corresponding to the relative astrometric uncertainty between optical and X-ray images: it is 0\farcs3 for S1 and S2, and 0\farcs5 for S3 (because L14-183 = S3 is a faint X-ray source several arcminutes from the ACIS aimpoint). The white contours show the combined 5.5-GHz and 9-GHz ATCA radio emission; more specifically, contour levels represent flux densities of $2^{n/2}$ times the local rms noise level, with $n =$ 3, 4, 5, etc. The cyan ellipses represent the ATCA beam for the stacked 5.5-GHz plus 9-GHz data: major axes are 1\farcs45\,$\times$\,0\farcs77, and the Position Angle is 1.2\degr. In the top left panel, the optical SNR B12-42 \citep{2012ApJS..203....8B} is also visible, $\approx$4$^{\prime\prime}$ to the south-west of S1; in the middle left panel, the optical and radio SNR B12-91 is visible, $\approx$4$^{\prime\prime}$ to the south-west of S2.
}

%
    \label{chand_img1}
    \vspace{0.5cm}
\end{figure*}


\begin{figure*}
    \centering
    \includegraphics[width=0.495\textwidth]{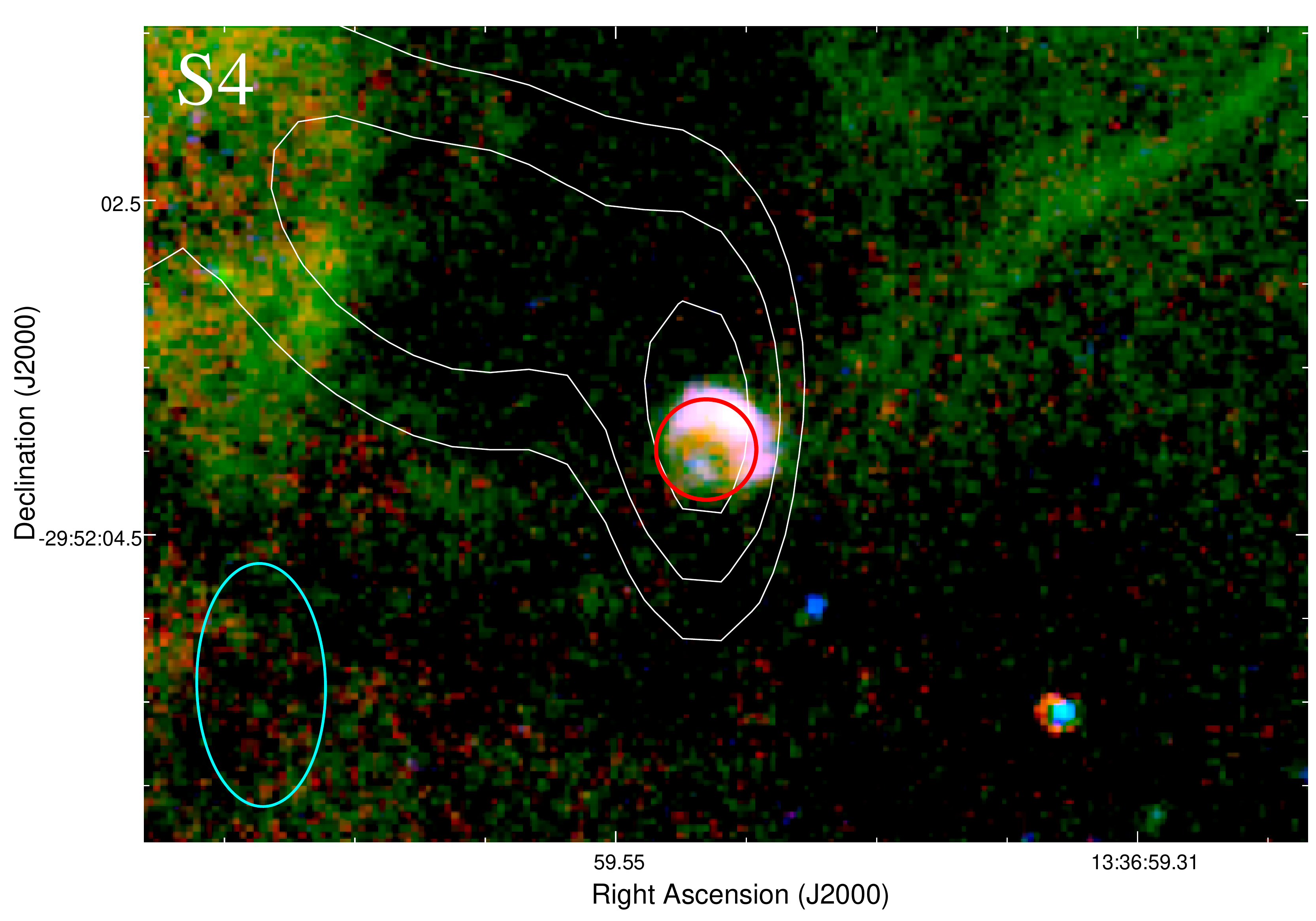}
    \includegraphics[width=0.495\textwidth]{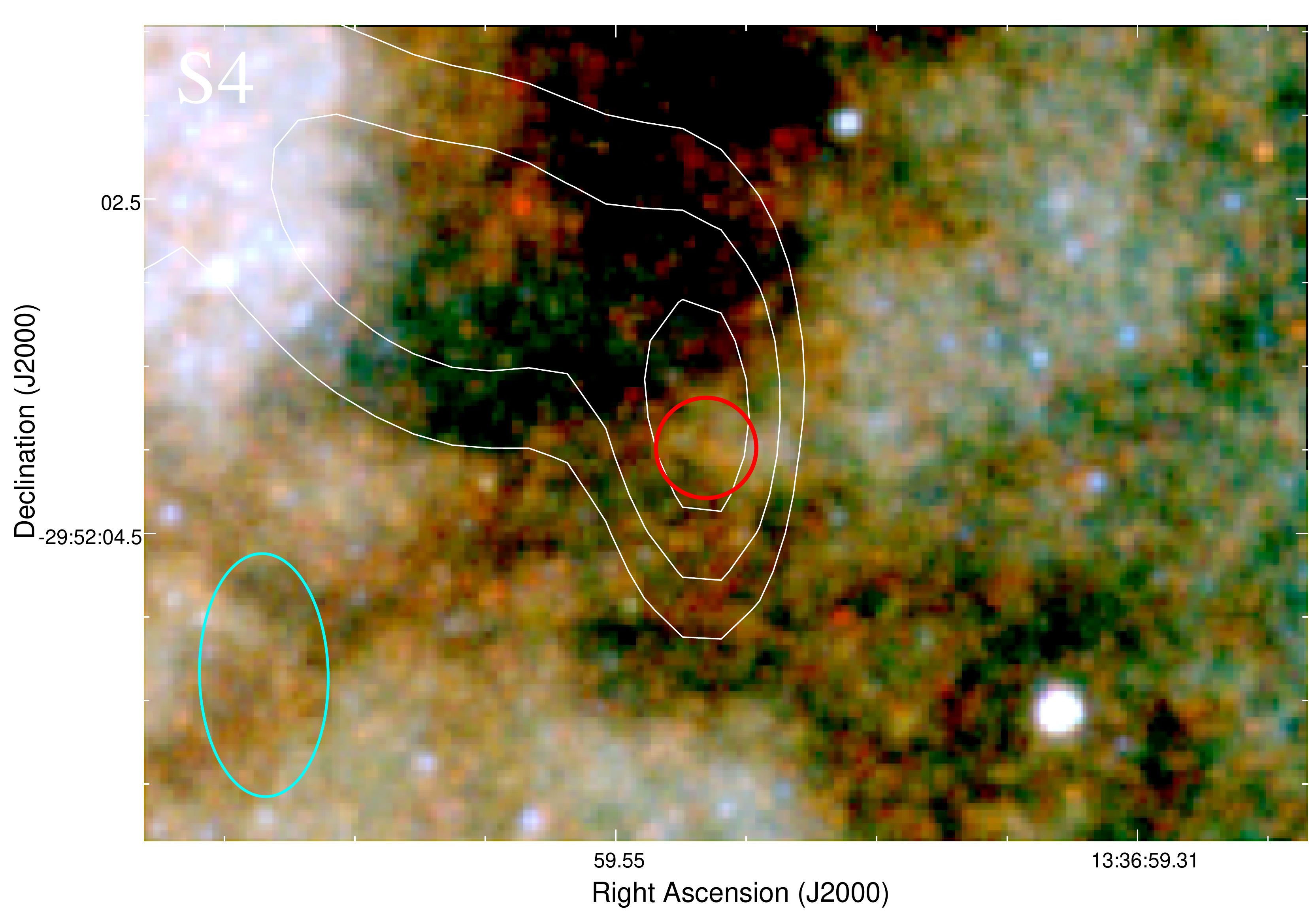}
    \\[-3pt]
    \includegraphics[width=0.495\textwidth]{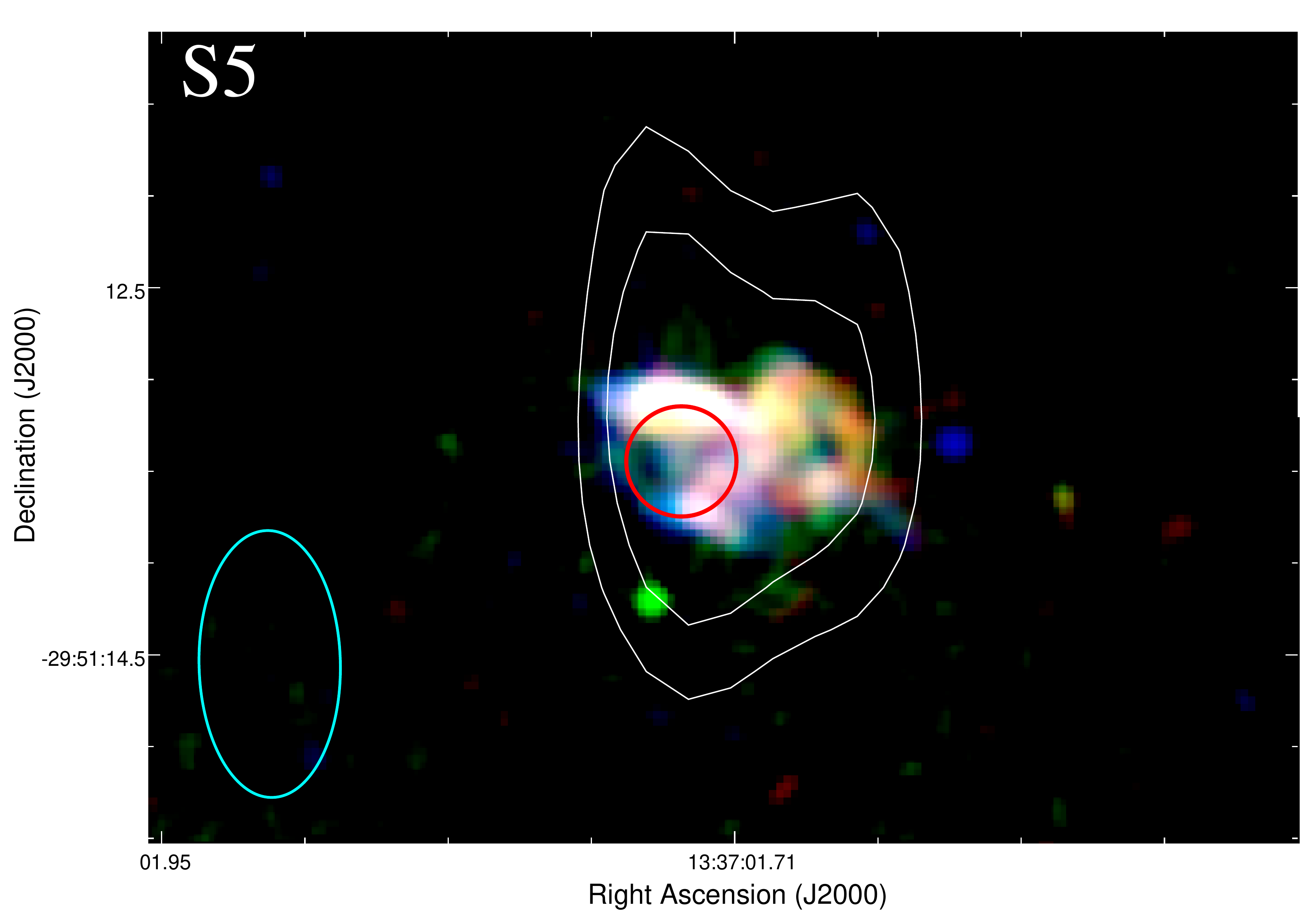}
    \includegraphics[width=0.495\textwidth]{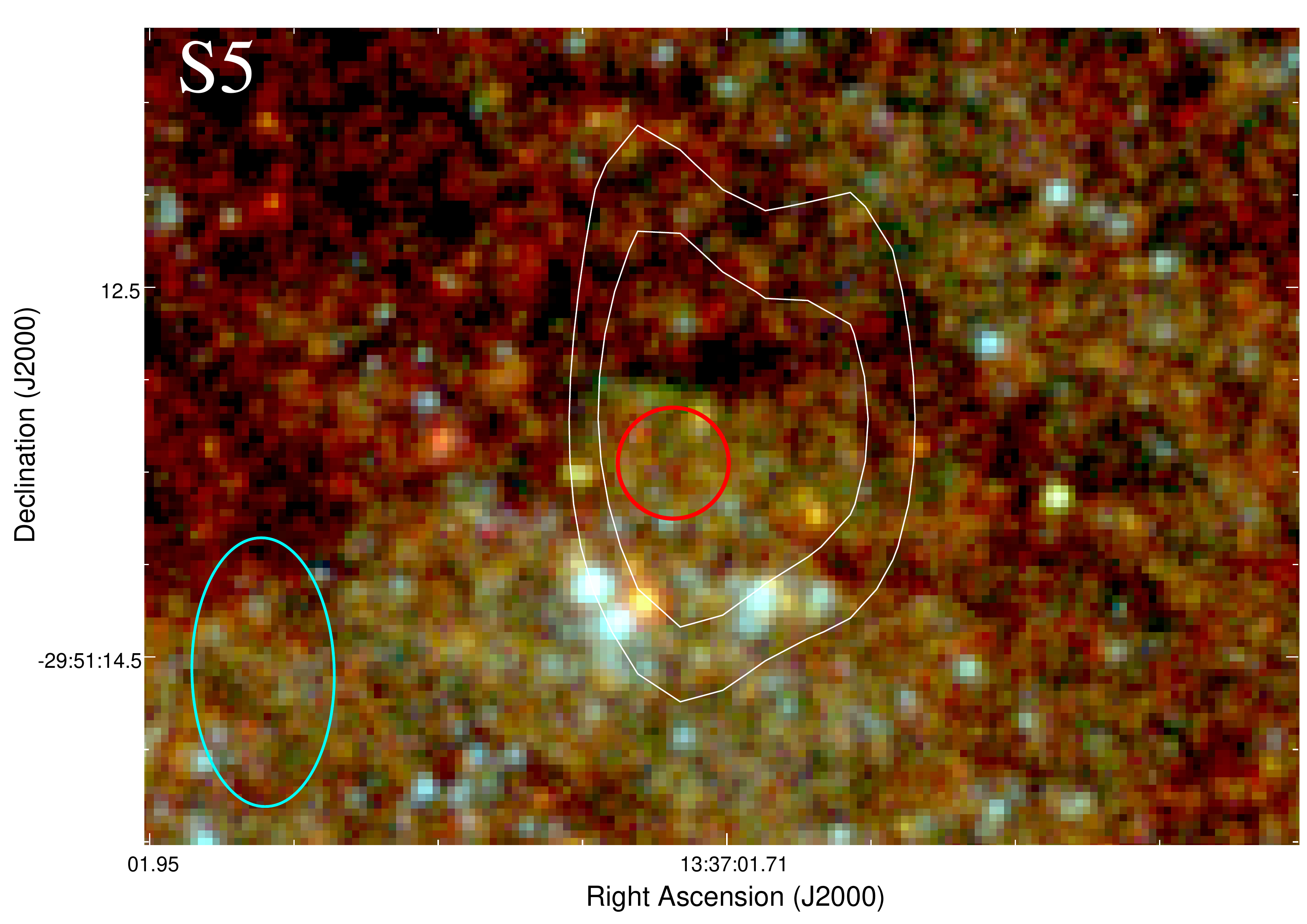}
    \\[-3pt]
    \includegraphics[width=0.495\textwidth]{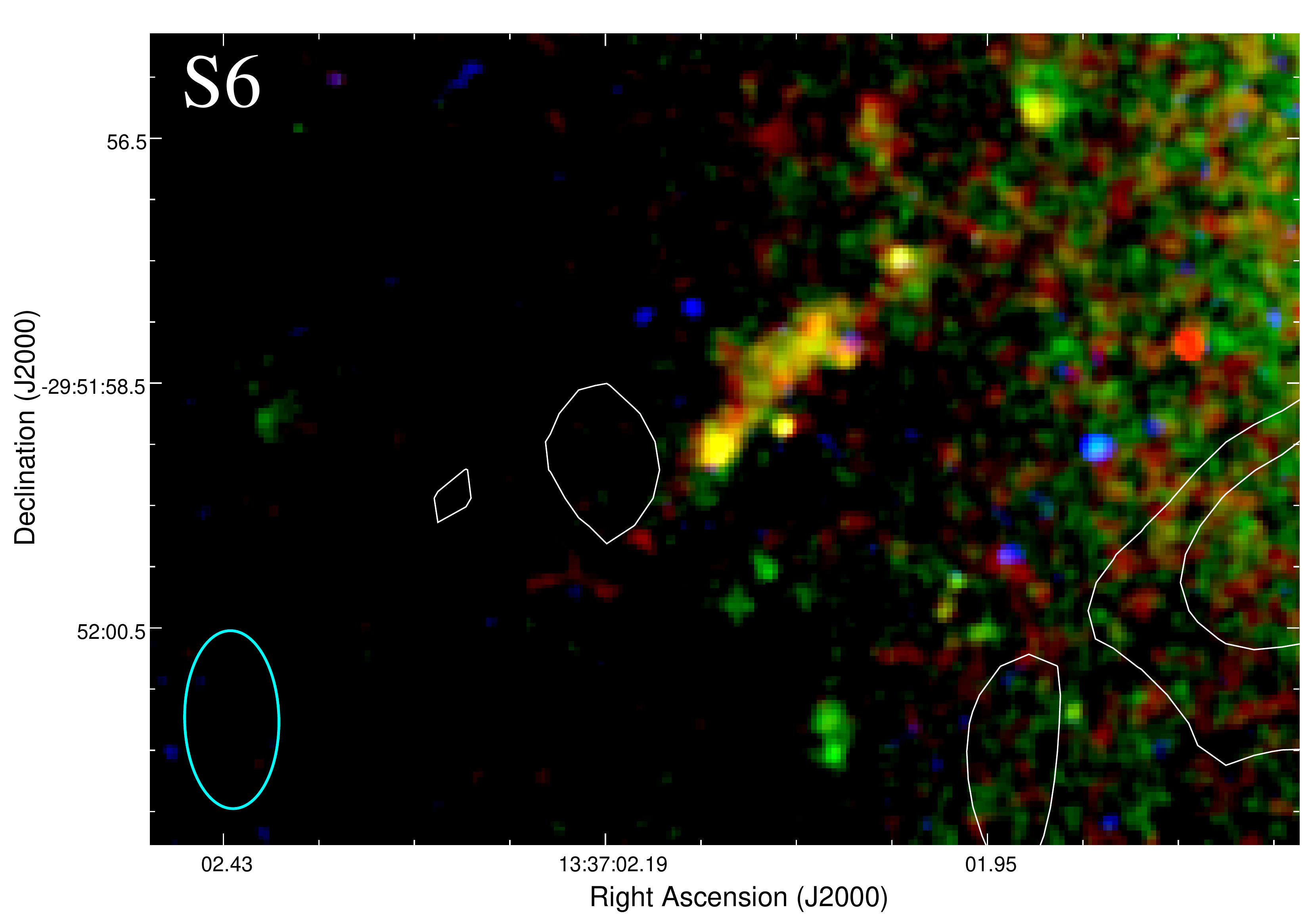}
    \includegraphics[width=0.495\textwidth]{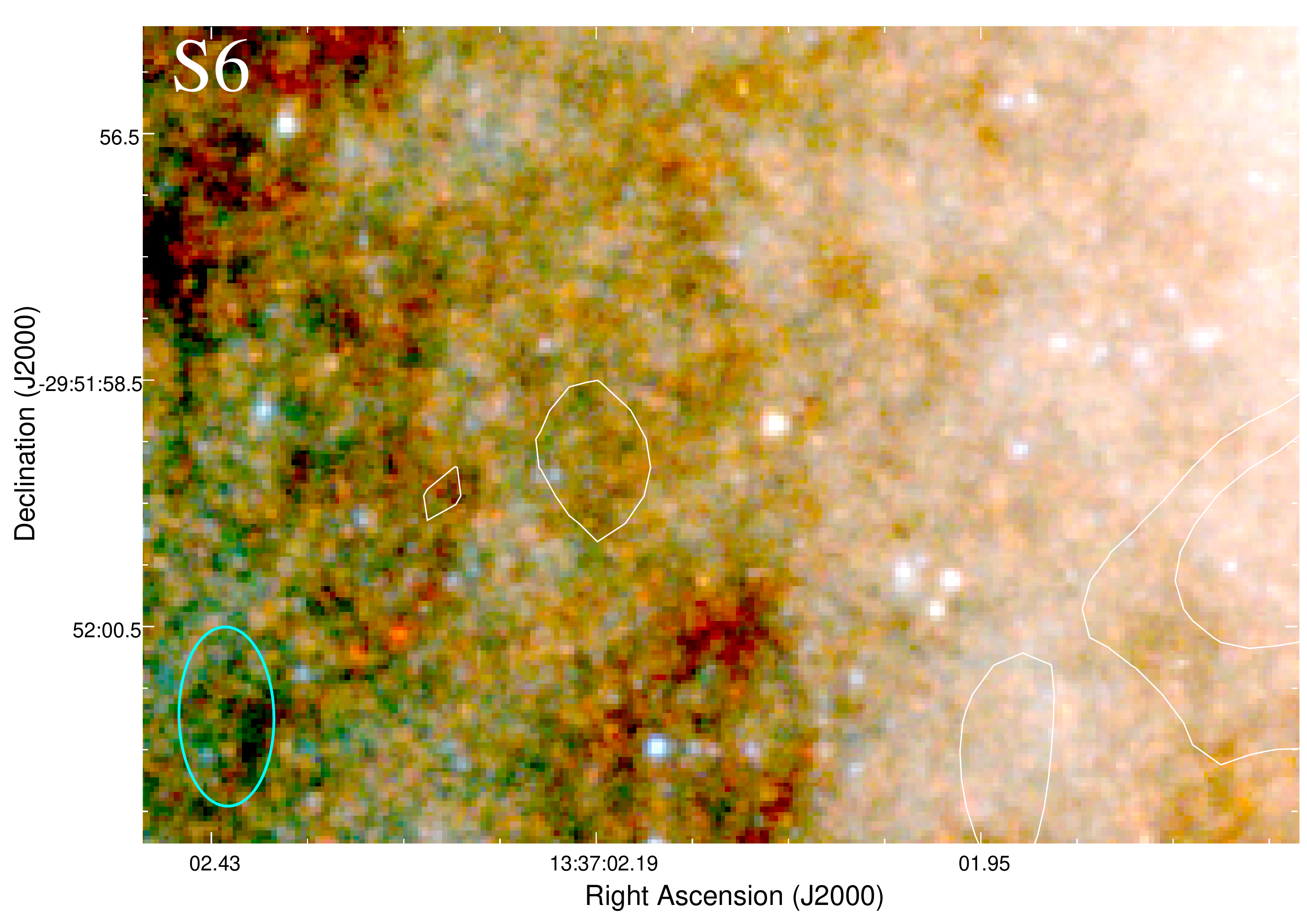}
    \vspace{-0.2cm}
\caption{Same as Figure 2, but for candidate SNRs S4, S5, and S6.  
For these three sources (located in Field 1 of our {\it HST} survey: \citealt{2014ApJ...788...55B}), the green colour in the right-hand panels corresponds to the F555W filter (not F547M); red is F814W and blue is F438W.
The  radius of the red circles (positional uncertainty of the corresponding {\it Chandra} sources) is 0\farcs3 for both S4 and S5; instead, S6 does not have a {\it Chandra} counterpart.
}
\label{chand_img2}
    \vspace{0.5cm}
\end{figure*}

\begin{figure*}
    \centering 
\includegraphics[width=0.495\textwidth]{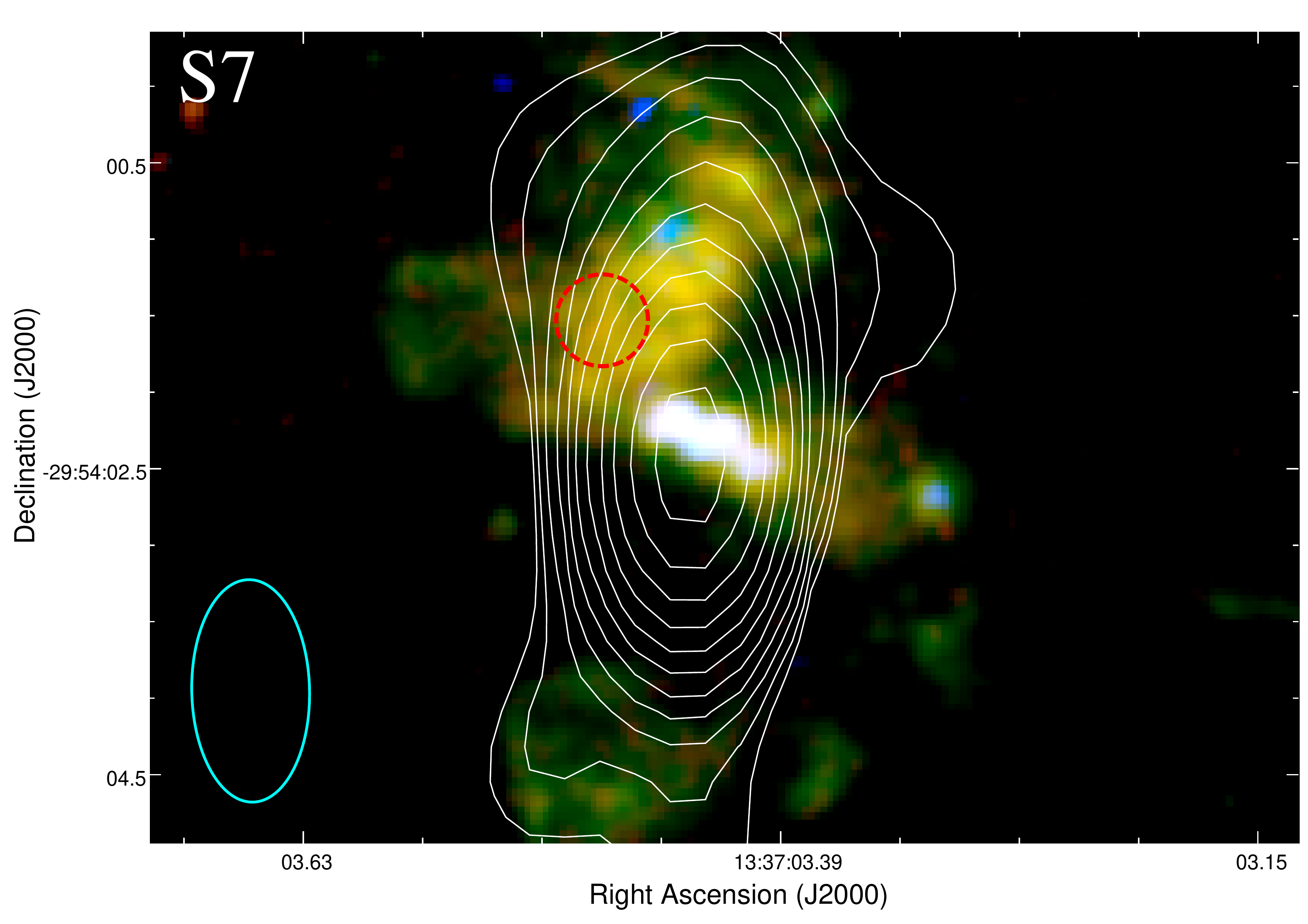}
\includegraphics[width=0.495\textwidth]{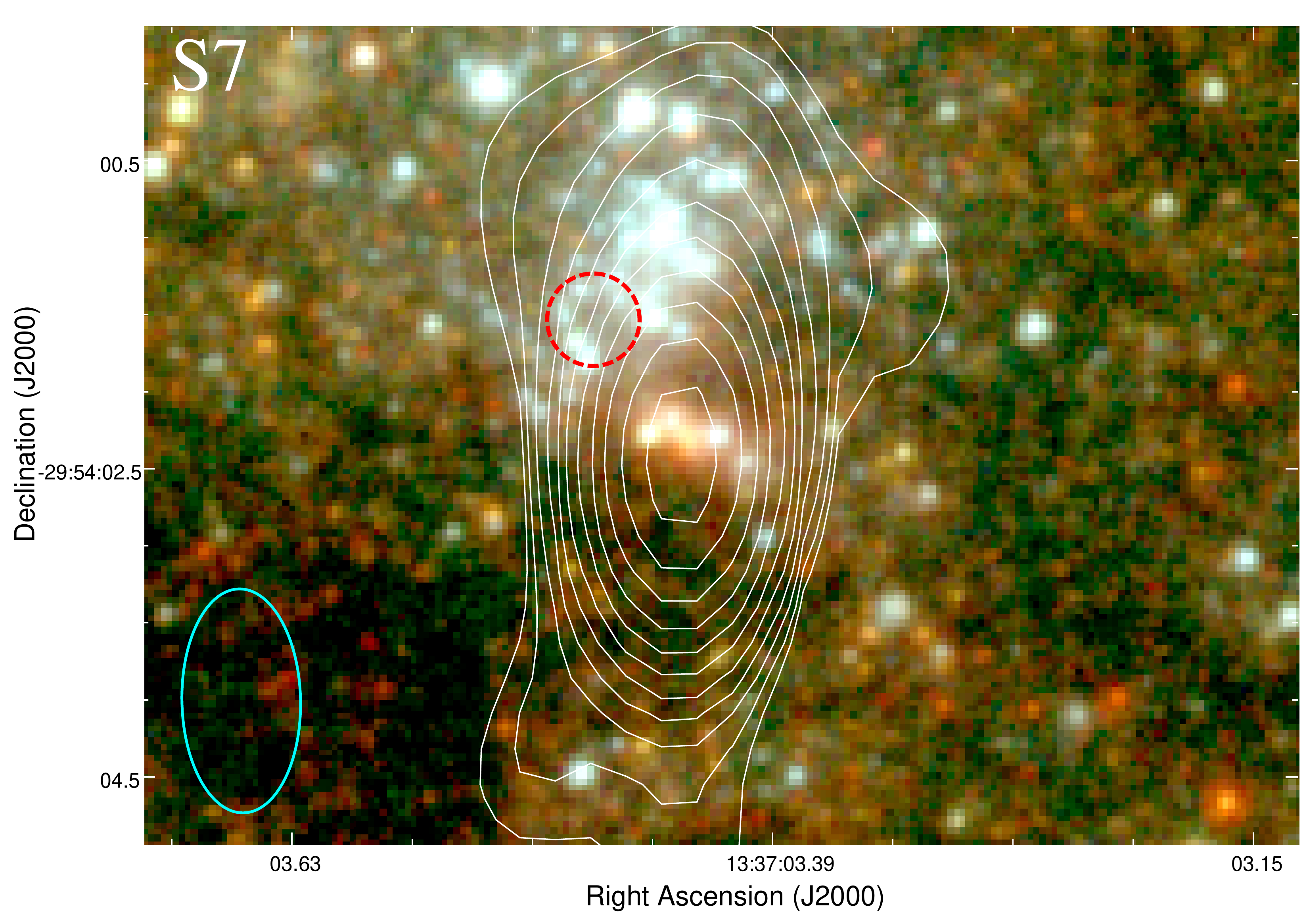}
\\[-3pt]
\includegraphics[width=0.495\textwidth]{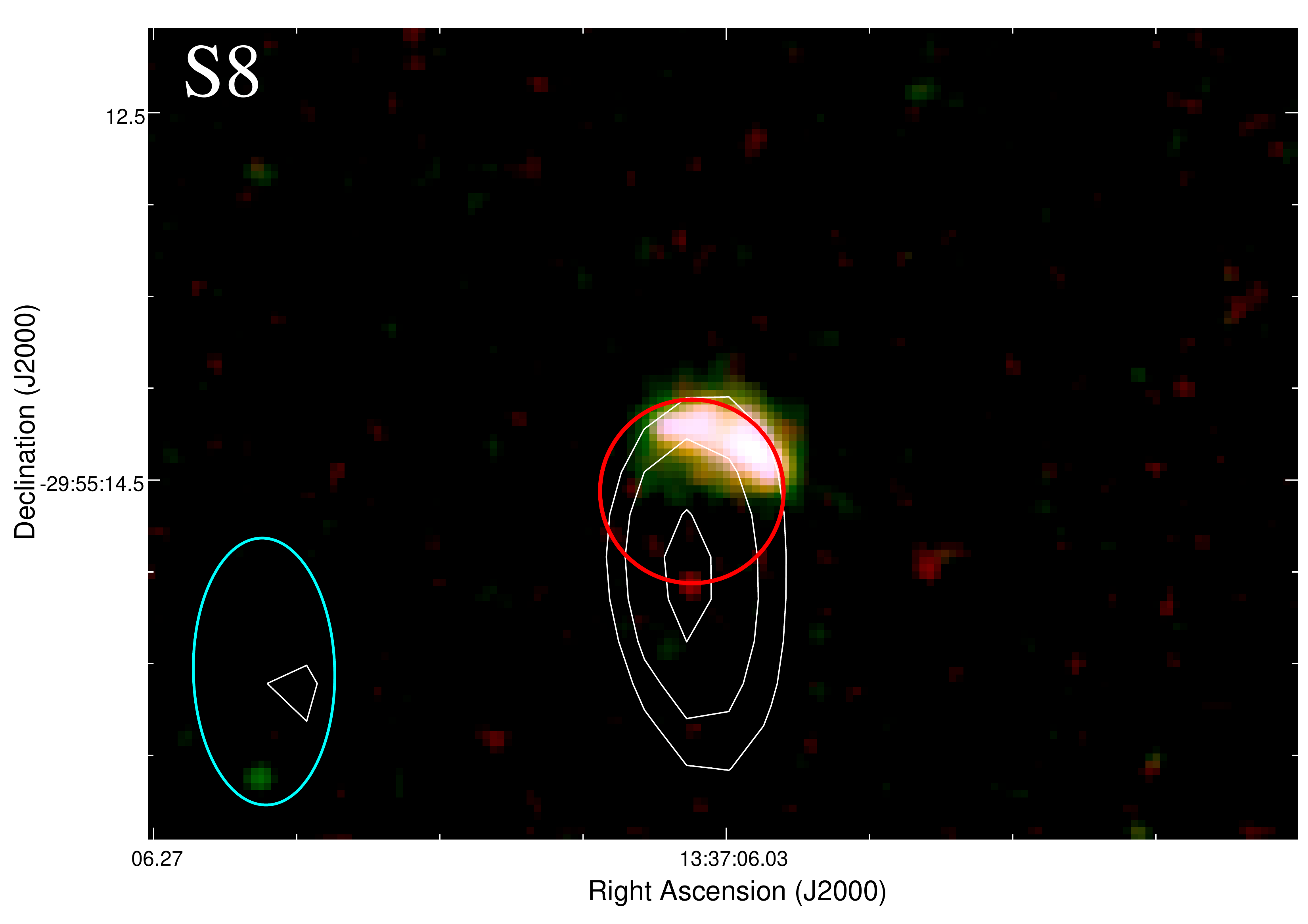}
\includegraphics[width=0.495\textwidth]{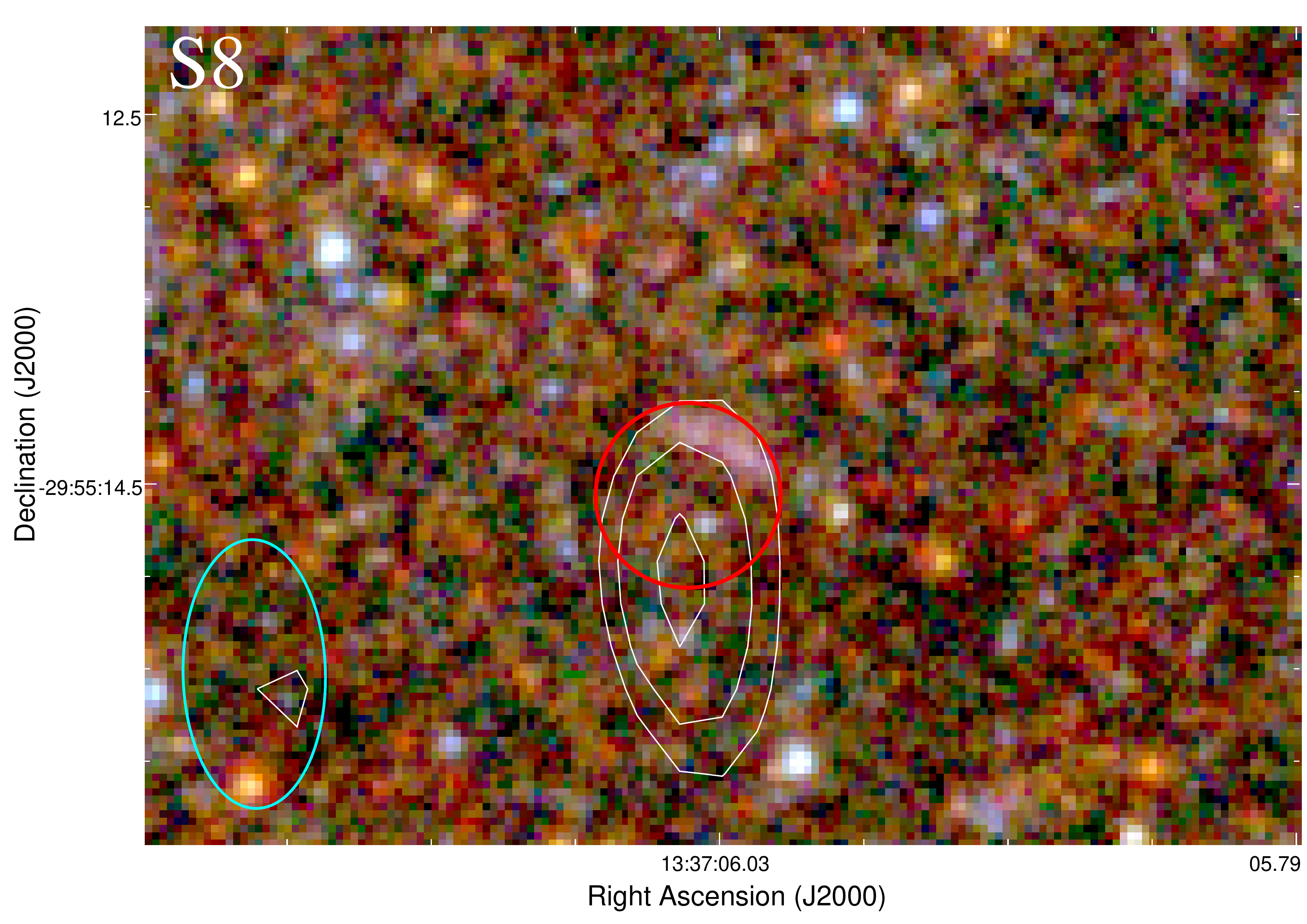}
\\[-3pt]
\includegraphics[width=0.495\textwidth]{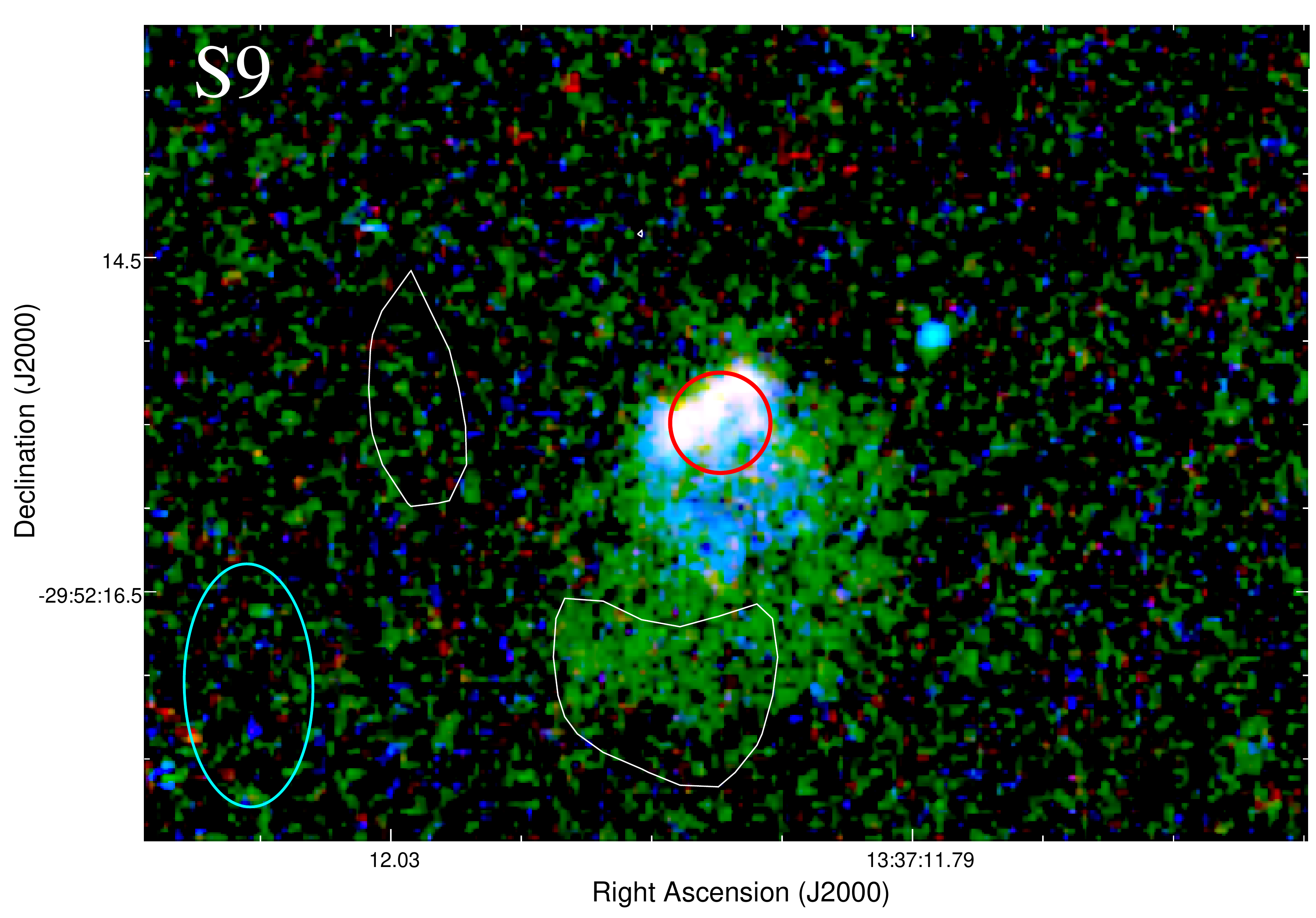}
\includegraphics[width=0.495\textwidth]{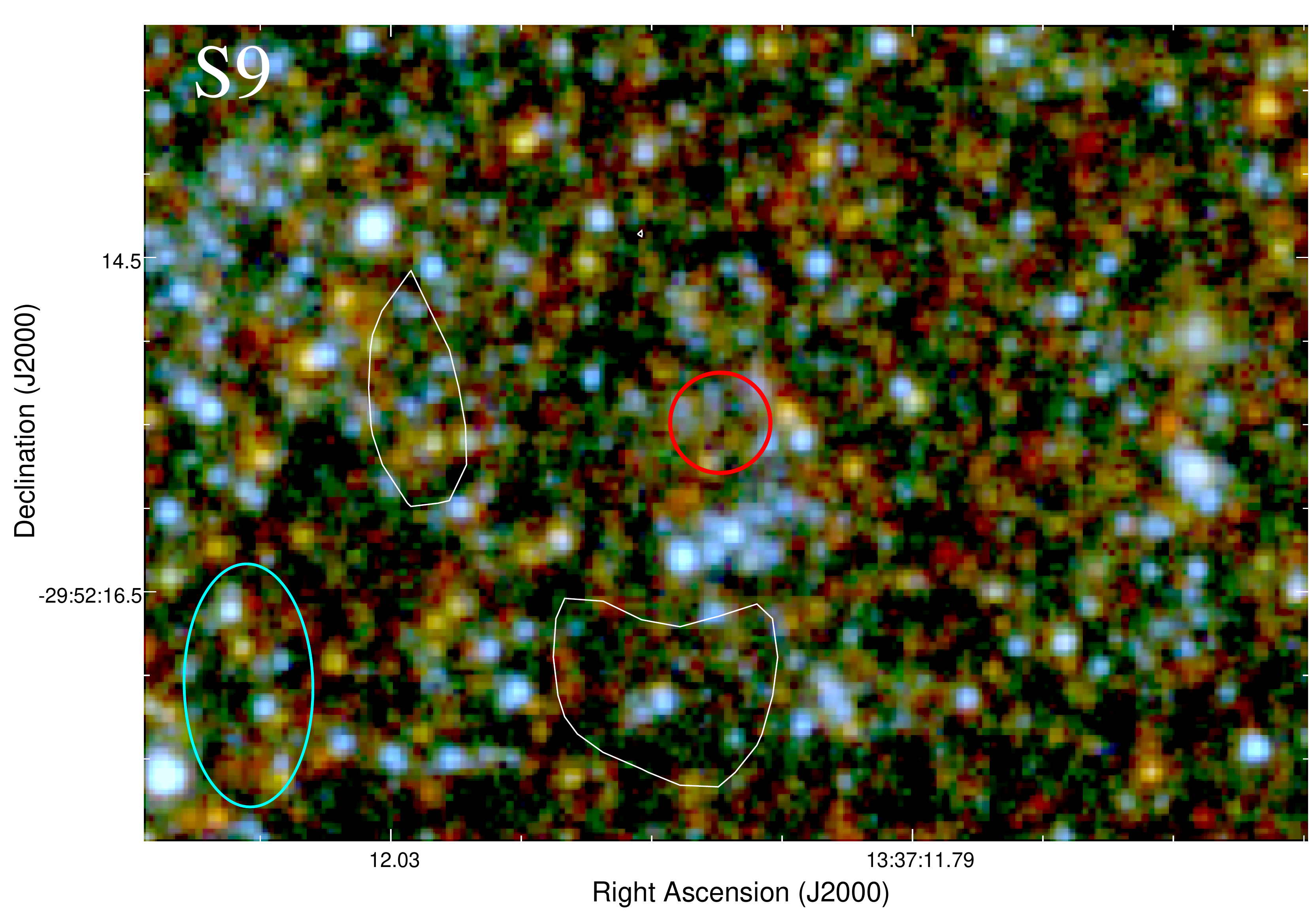}
\vspace{-0.2cm}
\caption{Same as Figure 2, but for candidate SNRs S7, S8, and S9.  
In the right panels, red is F814W, green is F547M, blue is F438W.
S7 has a {\it Chandra} source nearby (positional uncertainty of 0\farcs3) but we argue that it is not directly associated with the strong radio source: this is why the red circle is dashed. S8 does have an associated {\it Chandra} source, with a positional uncertainty (radius of the red circle) of 0\farcs5, because it was observed farther from the ACIS-S3 aimpoint. For S9, the radius of the {\it Chandra} error circle is 0\farcs3.
}
\label{chand_img3}
    \vspace{0.5cm}
\end{figure*}

\begin{figure}
    \centering
  \includegraphics[width=0.49\textwidth]{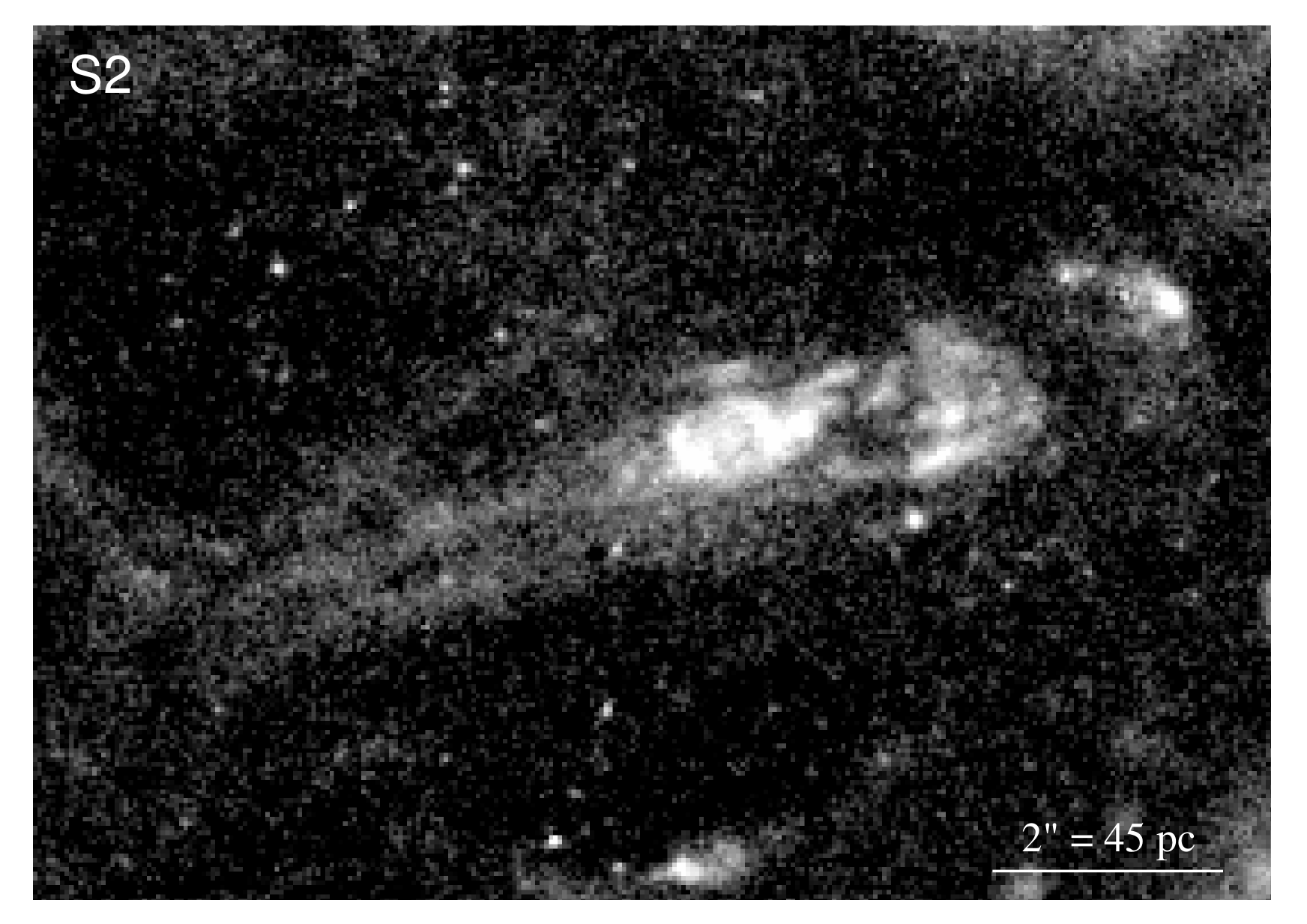}\\[-7pt]
   \includegraphics[width=0.49\textwidth]{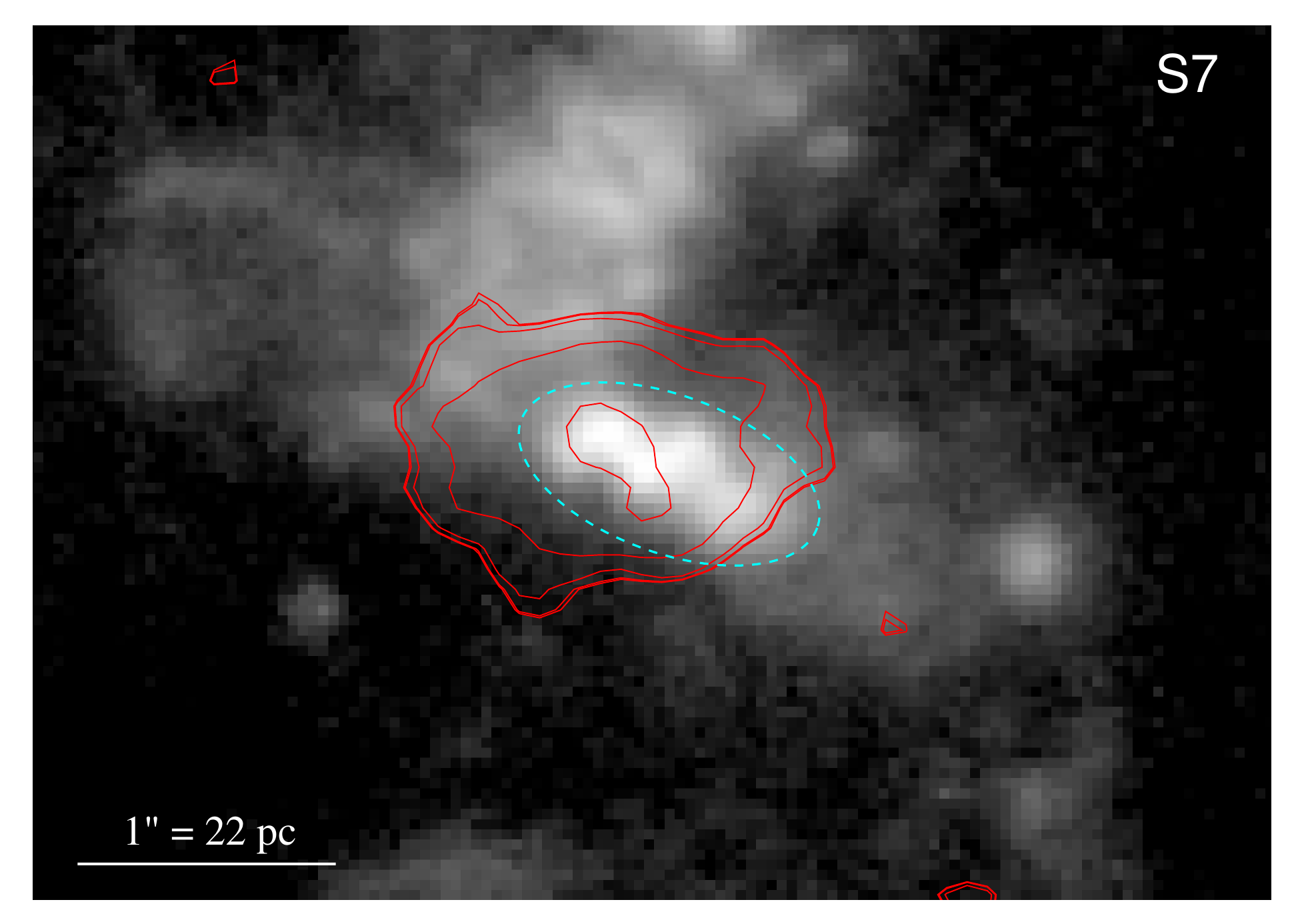}
\caption{Top panel: continuum-subtracted F657N (H$\alpha$ plus [N {\footnotesize{II}}) image of S2, highlighting the long linear structure with a possible bow shock at the western end; no [Fe II] image is available for this source. Bottom panel: continuum-subtracted F657N image of S7, with log-scale contours for the [Fe II] emission overplotted in red. The cyan ellipse shows the source extraction region used for the measurement of the line fluxes reported in Table 2. In both panels, north is up and east to the left, and a scale bar is shown.}
   \vspace{0.5cm}
\end{figure}

\begin{figure}
    \centering
\hspace{-0.5cm}    
\includegraphics[angle=270,width=0.8\textwidth]{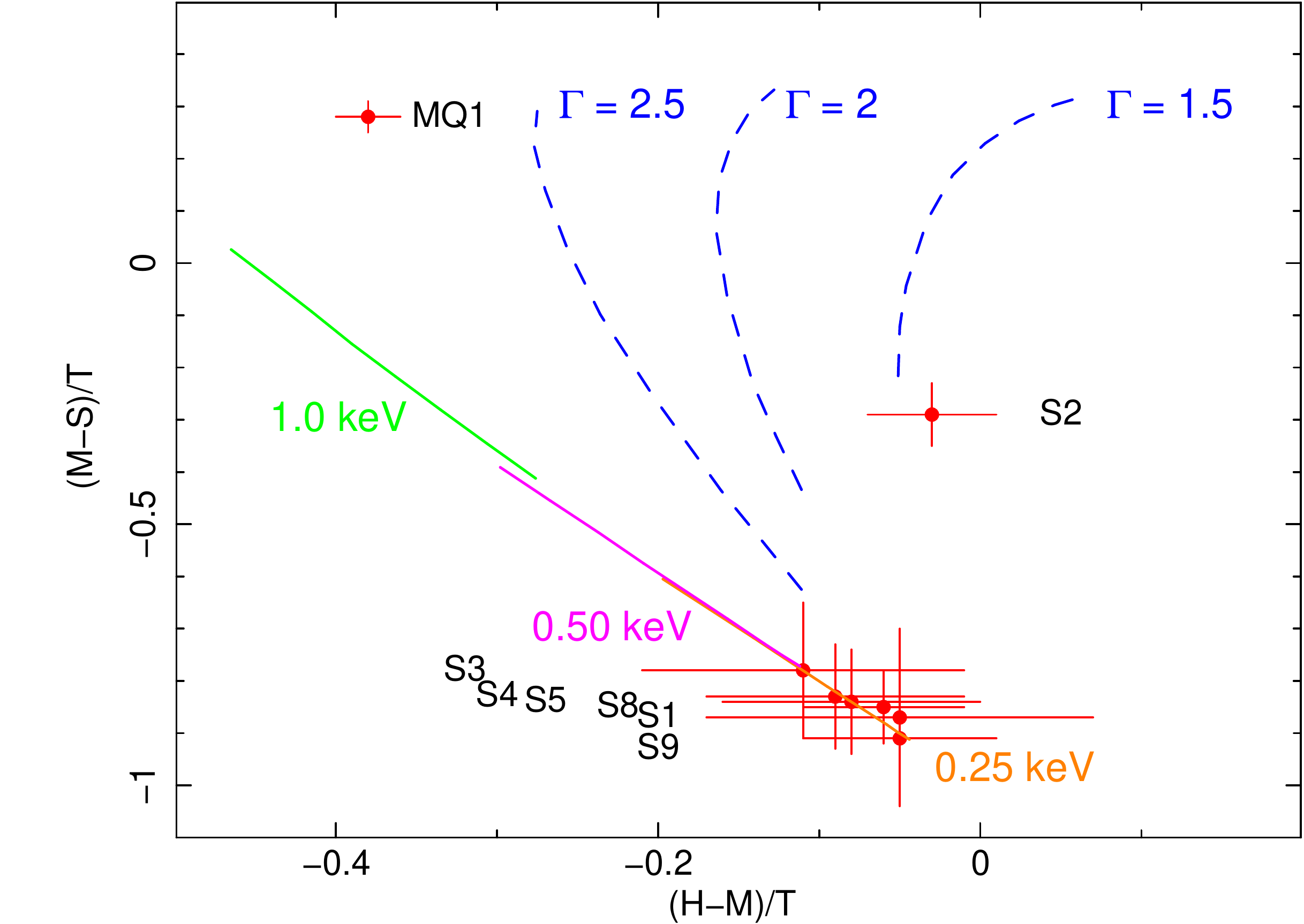}
\caption{{\it Chandra}/ACIS-S color-color plot for the seven point-like X-ray sources associated with our candidate SNRs, plus the MQ1 counterpart for comparison. The observed X-ray colors are defined as: S $=$ photon flux in the 0.35--1.1 keV band; M $=$ photon flux in the 1.1--2.6 keV band; H  $=$ photon flux in the 2.6--8.0 keV band; T $=$ photon flux in the 0.35--8.0 keV band. The color values are taken from \cite{2014ApJS..212...21L}. As a model grid, we overplotted the expected location of sources with power-law photon indices $\Gamma = 1.5$, 2, 2.5, and intrinsic absorption column running from $N_{\rm H} = 0$ to $N_{\rm H} = 7 \times 10^{21}$ cm$^{-2}$ (increasing from bottom to top). 
We also overplotted the expected colors of optically thin thermal plasma sources ({\it mekal} model) at temperatures $kT = 0.25$ keV, 0.50 keV and 1.0 keV, and $N_{\rm H}$ from 0 to $7 \times 10^{21}$ cm$^{-2}$ (increasing from let to right); this corresponds to the approximate temperature range of SNRs in the Large Magellanic Cloud \citep{2016A&A...585A.162M} and also to the empirical location of SNRs found in our previous studies of M\,83 \citep{2014ApJS..212...21L,2003A&A...410...53S}.
Based on X-ray colors alone, S2 = X139 is consistent with an X-ray binary in the hard state, S5 = X237 with an X-ray binary in the high/soft state, and the other counterparts are consistent with typical SNRs.}
   \vspace{0.5cm}
\end{figure}


\begin{figure*}
   \centering  
\includegraphics[angle=270,width=0.49\textwidth]{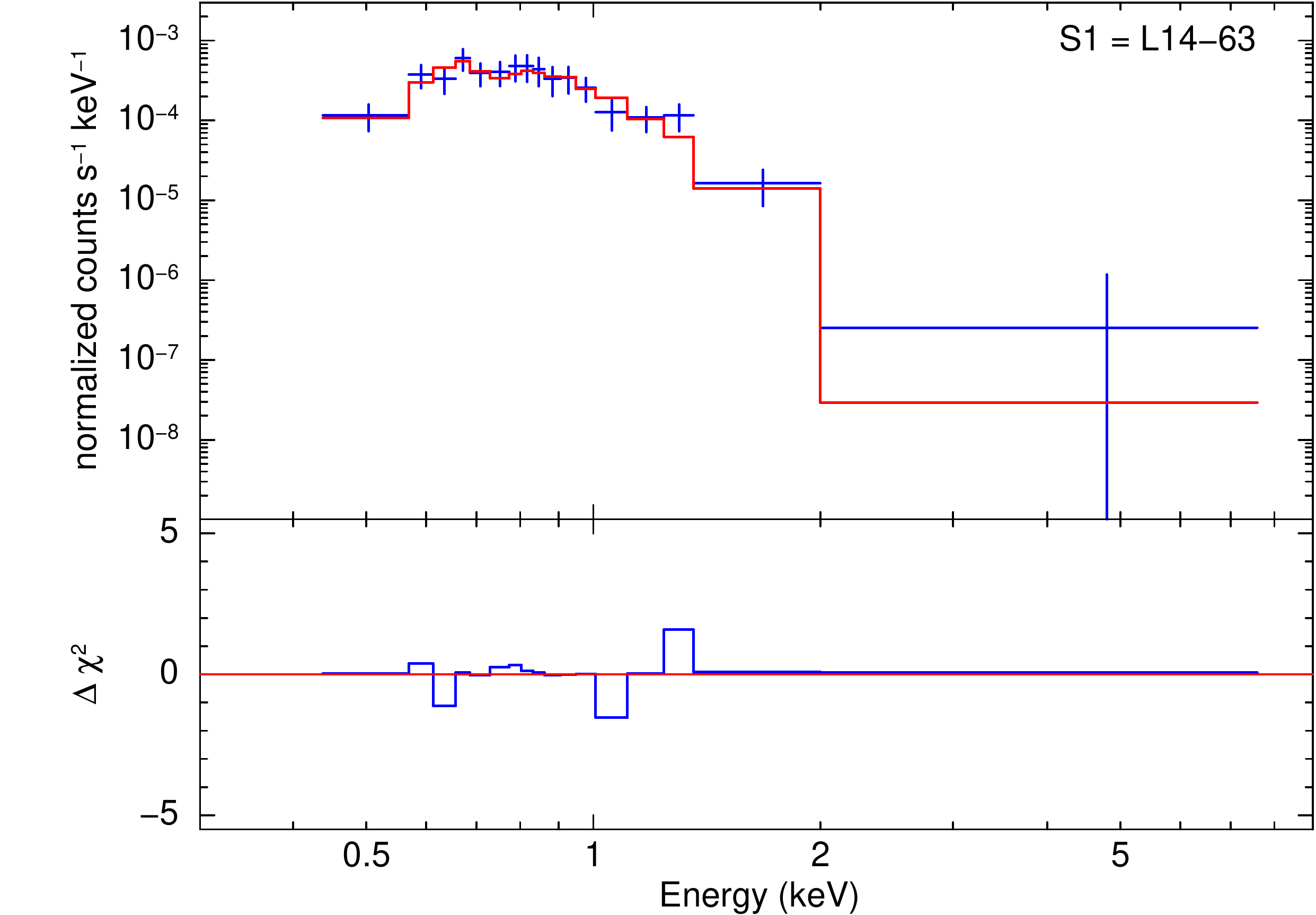}
\includegraphics[angle=270,width=0.49\textwidth]{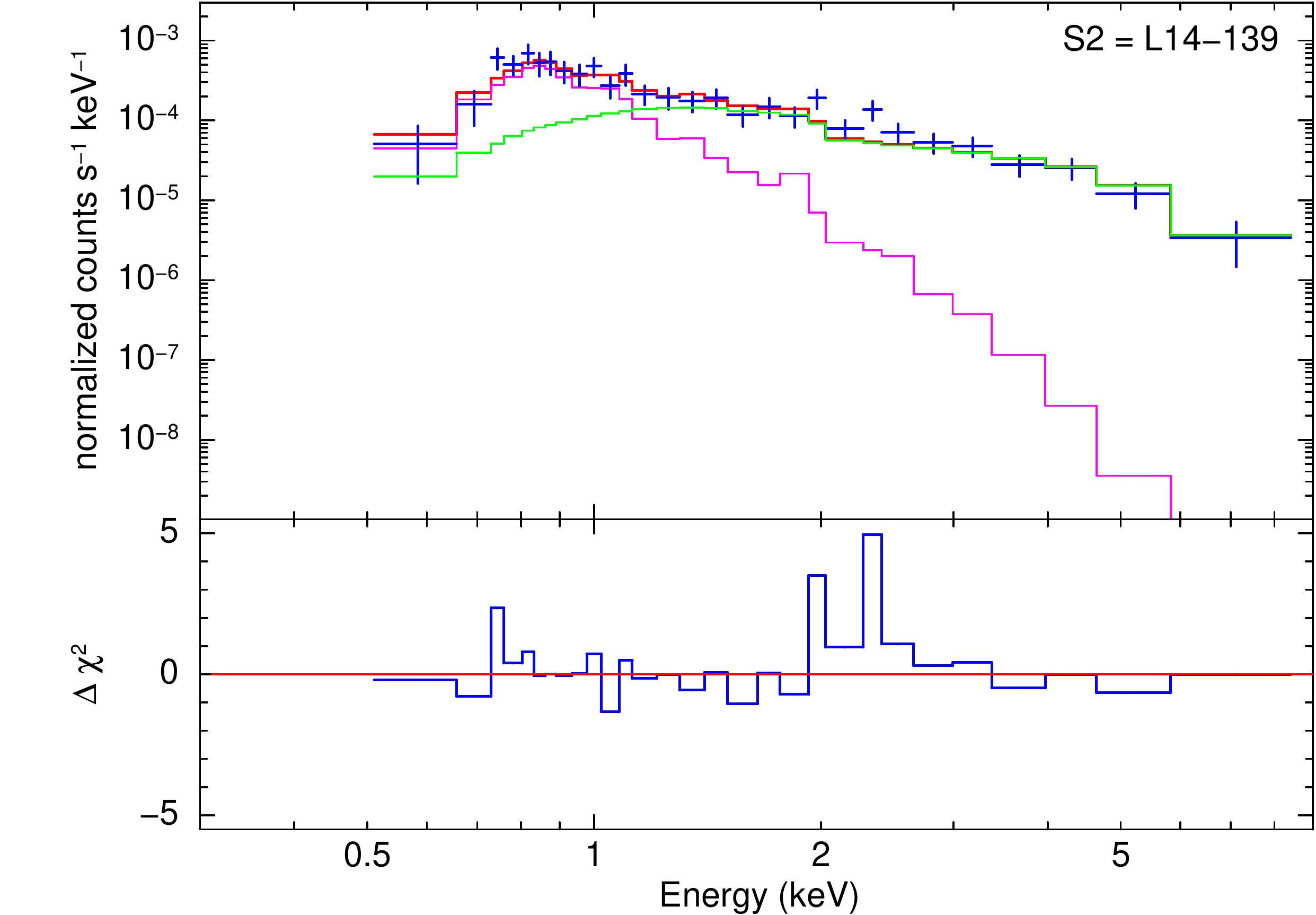}\\
\vspace{0.3cm}
\includegraphics[angle=270,width=0.49\textwidth]{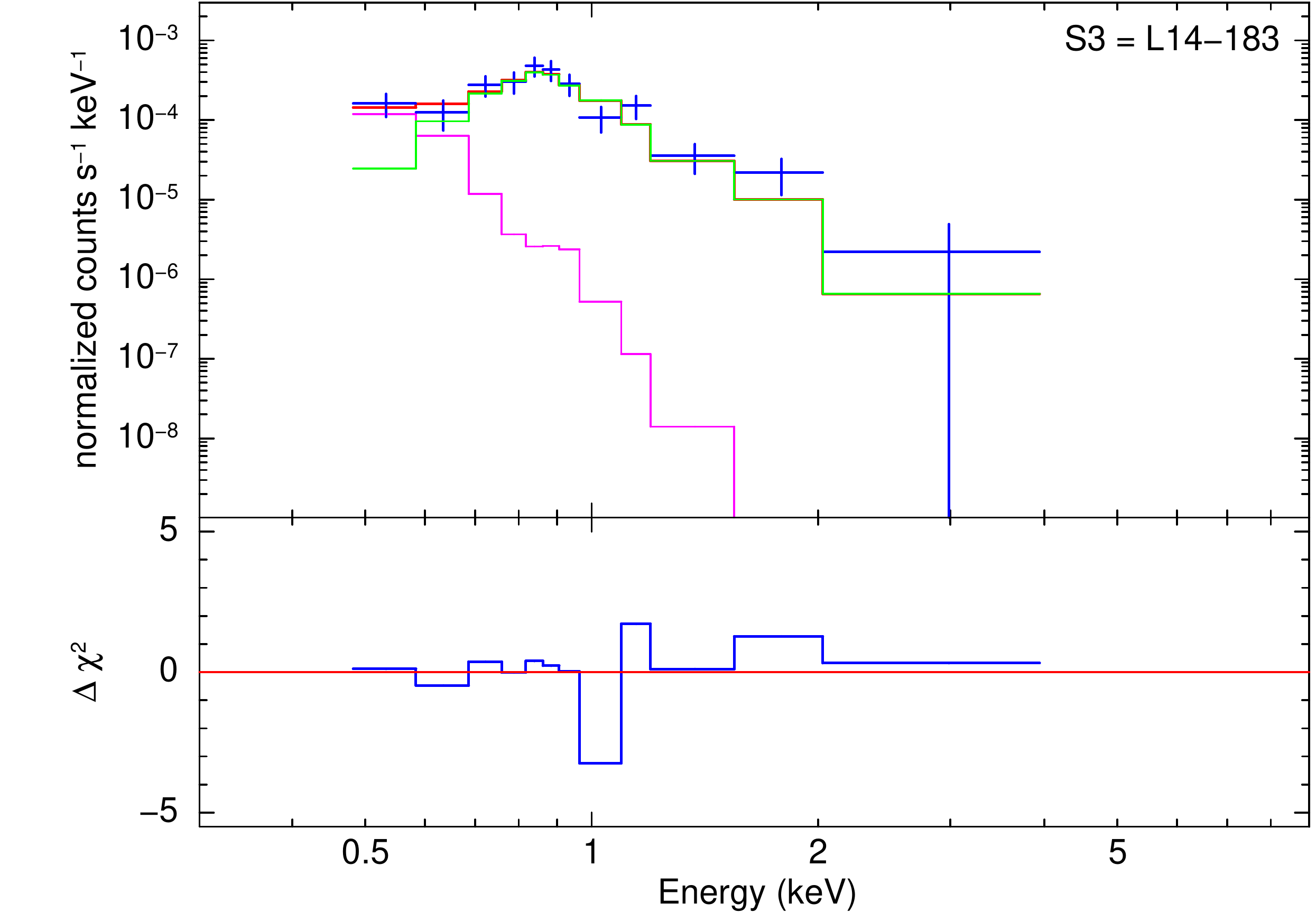}
\includegraphics[angle=270,width=0.49\textwidth]{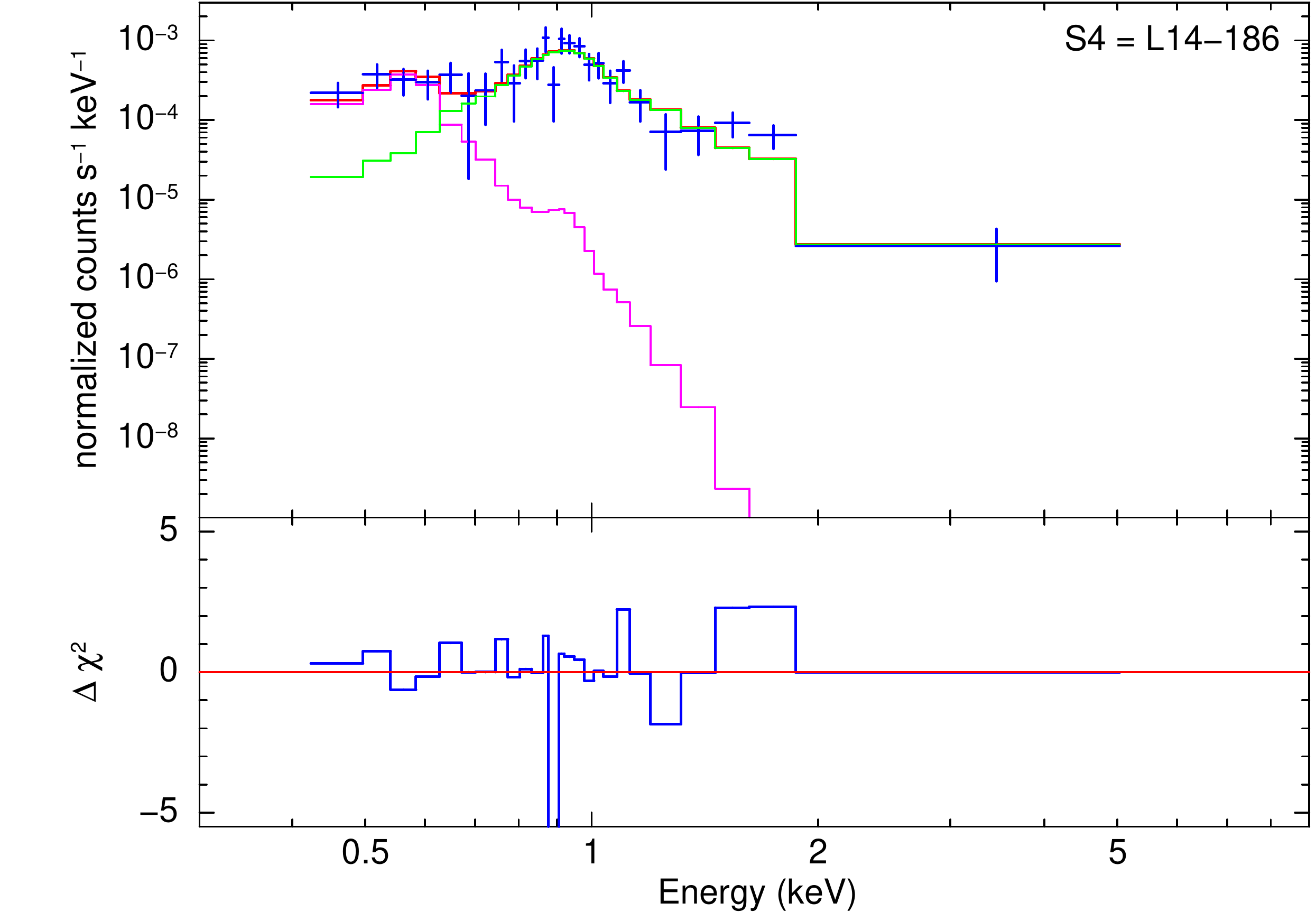}

\caption{Top left panel: stacked {\it Chandra} spectrum and $\chi^2$ residuals of X63 (X-ray counterpart of S1), fitted with a single-temperature thermal plasma model. Top right panel: stacked spectrum and $\chi^2$ residuals of X139 (X-ray counterpart of S2), fitted with a power-law model (photon index $\Gamma = 1.3 \pm 0.3$) plus a residual thermal-plasma component at $kT \approx 0.5$ keV. Bottom left panel: stacked spectrum and $\chi^2$ residuals of X183 (X-ray counterpart of S3), fitted with a two-temperature thermal plasma model. Bottom right panel: stacked spectrum and $\chi^2$ residuals of X186 (X-ray counterpart of S4), fitted with a two-temperature thermal plasma model. See Table 5 for the best-fitting parameters and corresponding fluxes and luminosities. In all panels, the red line is the combined best-fitting model; for two component models, the magenta line is the lower-temperature thermal plasma contribution, and the green line is the higher-temperature or power-law component. {\it Chandra}/ACIS spectral channels have been regrouped to a minimum of 15 counts per bin, for $\chi^2$ fitting.}

   \vspace{0.5cm}
\end{figure*}

\begin{figure}
    \centering
\hspace{-0.4cm}    
\includegraphics[angle=270,width=0.49\textwidth]{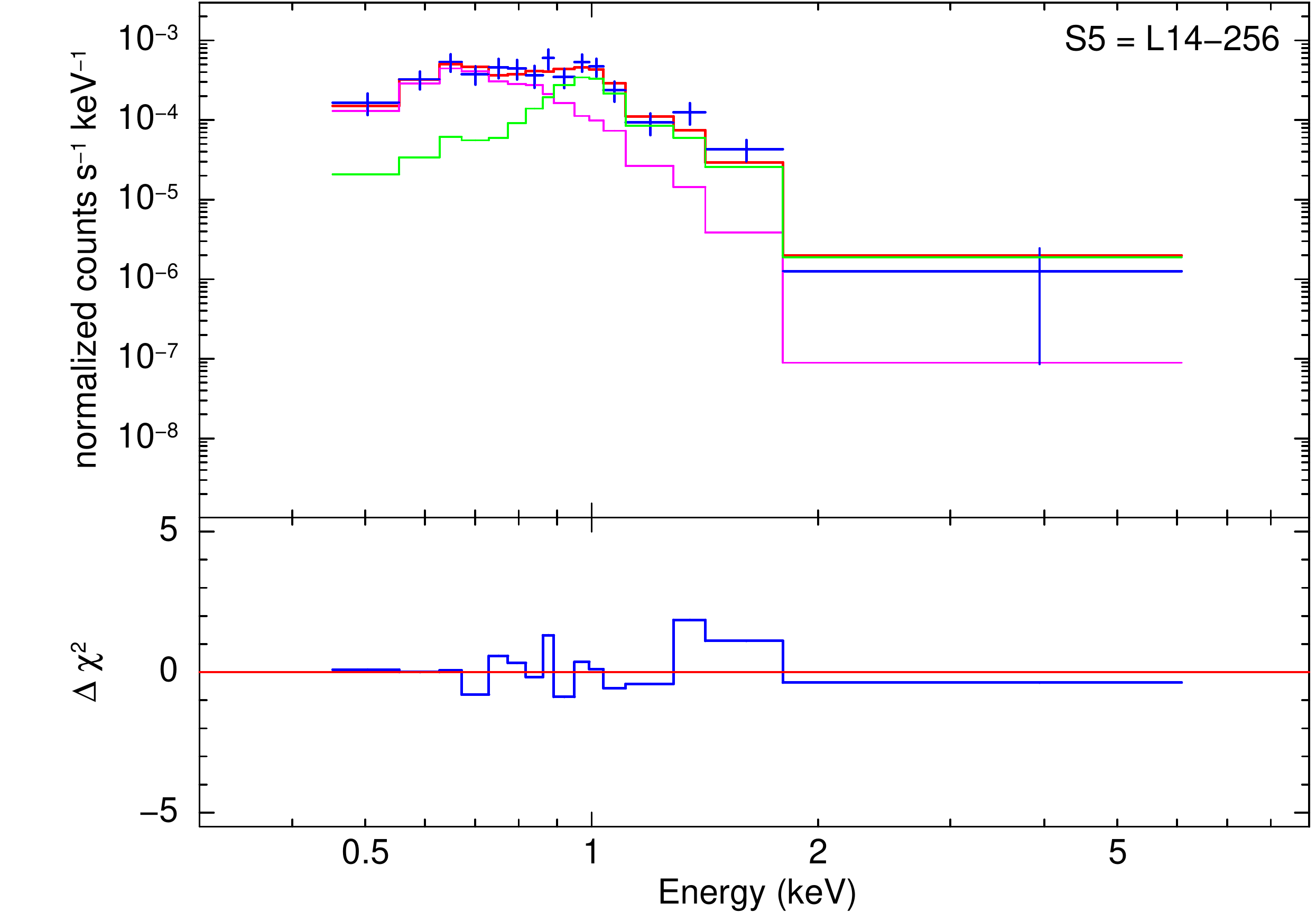}\\
\hspace{-0.4cm} 
\includegraphics[angle=270,width=0.49\textwidth]{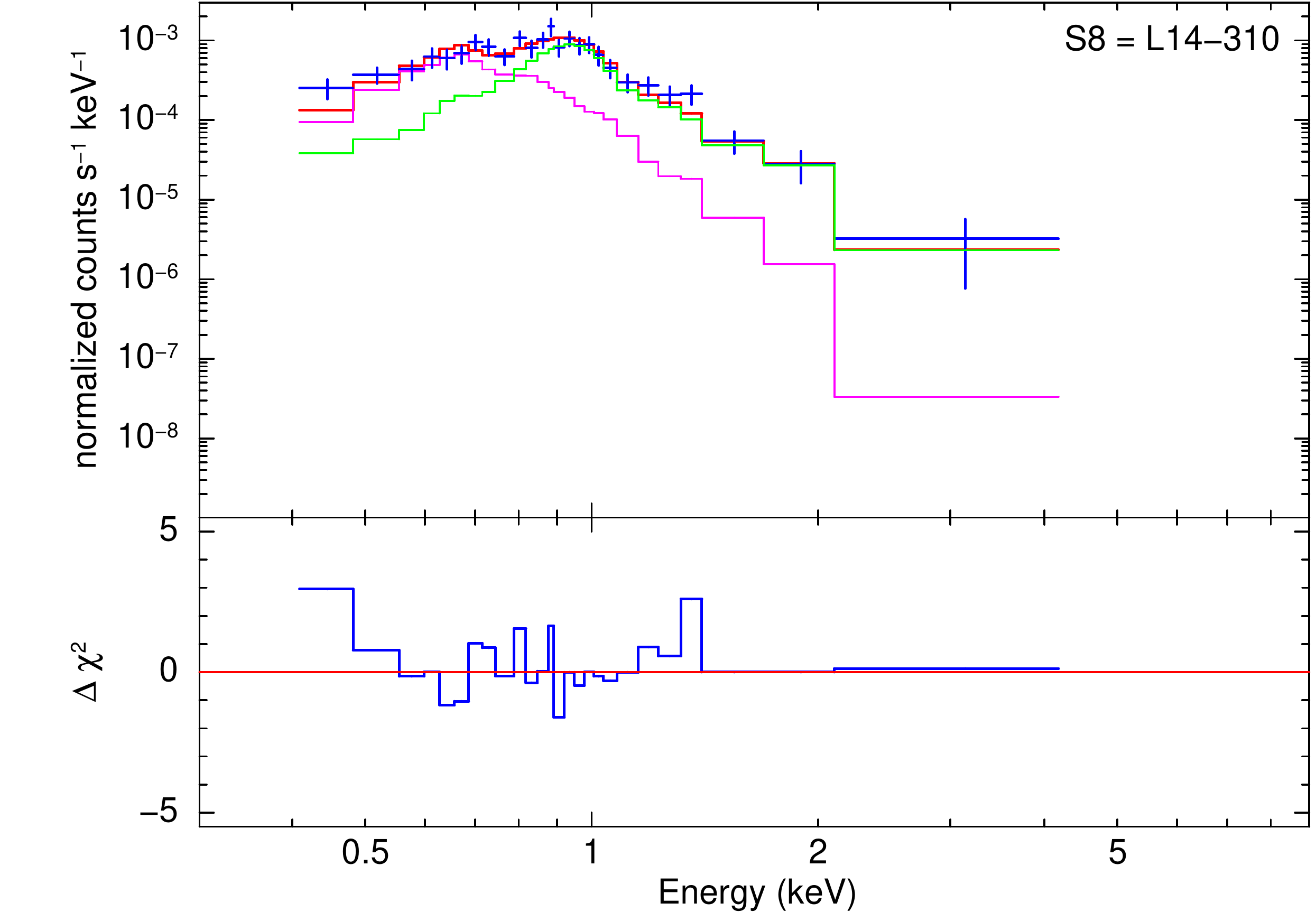}\\
\hspace{-0.4cm}        
\includegraphics[angle=270,width=0.49\textwidth]{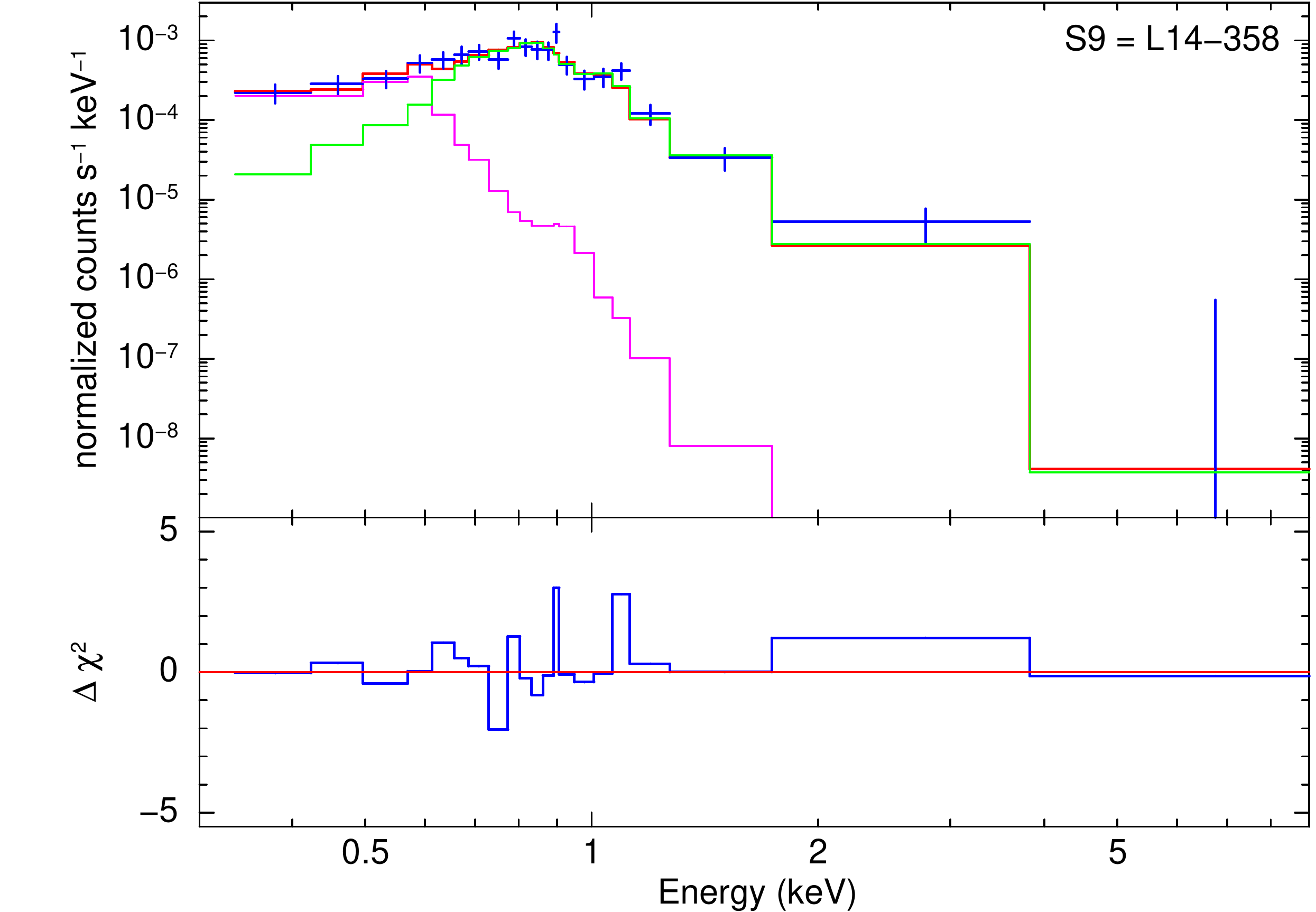}\\
\caption{Top panel: stacked spectrum and $\chi^2$ residuals of X256 (X-ray counterpart of S6), fitted with a two-temperature thermal plasma model. Middle panel: stacked spectrum and $\chi^2$ residuals of X310 (X-ray counterpart of S9), fitted with a two-temperature thermal plasma model. Bottom panel: stacked spectrum and $\chi^2$ residuals of X358 (X-ray counterpart of S10), fitted with a two-temperature thermal plasma model. See Table 5 for the best-fitting parameters and corresponding fluxes and luminosities. Red, magenta and green lines are defined as in Figure 7. The data have been regrouped to $>$15 counts per bin.}
   \vspace{0.5cm}
\end{figure}

\end{document}